\documentclass[12pt,aps,pra,amsmat,amssymb,amsfonts,superscriptaddress,onecolumn, notitlepage]{revtex4-2}
\usepackage[section]{placeins}
\usepackage{relsize}
\usepackage{amsfonts}
\usepackage{hyperref}
\usepackage{bbm}
\usepackage{mathtools}
\usepackage{tikz}
\usepackage{physics}
\usepackage{amsthm}
\usepackage{amsmath}
\usepackage{amssymb}
\usepackage{tensor}
\usepackage{bigints}
\usepackage{mdframed}
\usepackage{xcolor}
\usepackage{graphicx}
\usepackage{float}
\usepackage{booktabs}
\usepackage{cleveref}
\usepackage{setspace}
\usepackage{subcaption}
\usepackage{tensor}
\usepackage{orcidlink}
\usepackage{enumitem}

\newcommand{\qexp}[1]{\langle#1\rangle}

\begin{document}
\title{From Classical to Quantum Information Geometry: A Guide for Physicists}
\author{J. Lambert}
\email{lambej3@mcmaster.ca}
\author{E. S. S{\o}rensen\,\orcidlink{0000-0002-5956-1190}}
\email{sorensen@mcmaster.ca}
\affiliation{Department of Physics \& Astronomy, McMaster University
1280 Main St.\ W., Hamilton ON L8S 4M1, Canada}
\date{November 2022}

\begin{abstract}
Recently, there has been considerable interest in the application of information geometry to quantum many body physics. This interest has been driven by three separate lines of research, which can all be understood as different facets of quantum information geometry. First, the study of topological phases of matter characterized by Chern number is rooted in the symplectic structure of the quantum state space, known in the physics literature as Berry curvature. Second, in the study of quantum phase transitions, the fidelity susceptibility has gained prominence as a universal probe of quantum criticality, even for systems that lack an obviously discernible order parameter. Finally, the study of quantum Fisher information (QFI) in many body systems has seen a surge of interest due to its role as a witness of genuine multipartite entanglement and owing to its utility as a quantifier of quantum resources, in particular those useful in quantum sensing. Rather than a thorough review, our aim is to connect key results within a common conceptual framework that may serve as an introductory guide to the extensive breadth of applications, and deep mathematical roots, of quantum information geometry, with an 
intended audience of researchers in quantum many body and condensed matter physics.
\end{abstract}
\maketitle
\newpage
\tableofcontents
\newpage

\section{Overview and Motivations}\label{sec:Introduction}

The importance of geometry to the study of quantum many body systems has emerged in several major contexts. First, in the groundbreaking work of Thouless and collaborators~\cite{thouless1982quantized}, who noticed the relationship between the geometric phase~\cite{pancharatnam1956generalized,berry1984quantal} of a single electron wave function, and the topology of the band structure in two-dimensional insulators. Second, in the work of Zanardi~\cite{zanardi2006ground}, who applied the notion of state space geometry to the study of quantum phase transitions, and termed the resulting geometric quantity \emph{fidelity susceptibility}~\cite{gu2010fidelity}. The fidelity susceptibility and geometric phase are two halves of a deeper whole. The former defines the Riemannian structure of the state space~\cite{provost1980riemannian} while the latter defines the symplectic structure. Taken together, the study of these two structures is the subject of quantum information geometry~\cite{brody2001geometric}. The word information here arises due to the relationship between the geometry of the state space, and the degree to which the parameters that define the location of a state in state space may be estimated in a fixed number of measurements. In classical statistics this is the concept of Fisher information~\cite{fisher1925theory}, which was generalized to the quantum context for pure states by Wootters~\cite{wootters1981statistical}, and then to the case of mixed states by Braunstein and Caves~\cite{braunstein1994statistical}. As we shall see, the relationship between geometry and information has important implications for a number of fields, including quantum many body physics, quantum sensing and quantum computing.  

Many of the topics touched upon in the present guide have  been the subject of a dedicated review. For example, for a review of fidelity susceptibility one can consult~\cite{gu2010fidelity} or more recently~\cite{carollo2020geometry}, and for a review of the applications of the geometric phase to condensed matter systems one might consult~\cite{bohm2003geometric}. A recent and very thorough review of quantum Fisher information in the context of metrology can be found in Ref.~\cite{liu2019quantum}. For a mathematically rigorous review of the classical Fisher information in the context of statistical physics, one can consult Ref.~\cite{brody2008vapourliquid}. In its bibliographical notes, this review also contains a very complete history of the field of information geometry from a mathematical perspective. Nonetheless, over the last decade a number of important connections have emerged between state space geometry, quantum sensing~\cite{giovannetti2006quantum,giovannetti2011advances,toth2014metrologyfrominformation,degen2017quantum}, multipartite entanglement~\cite{toth2012multipartite,hyllus2012fisher}, topology~\cite{kemp2021nested,klees2020microwave,klees2021ground}, and response functions~\cite{hauke2016measuring,shitara2016determining}, which demand a contemporary perspective, easily accessible to a quantum many-body audience. The present guide intends to serve two purposes. First, as a pedagogical introduction, using a unified notation and conceptual framework, to information geometry in quantum many body physics. 
Second, as a guide where key results in the field from the last two decades
are gathered in one place,
offering new researchers an opportunity to orient themselves in the field, and providing experienced researchers with an entry point for future inquiries. While we avoid rigorous proofs, we strive to summarize the key foundational ideas in the mathematics literature, providing references to the original research literature.

Much of the utility of quantum information geometry arises from the distinction between classical fluctuations and quantum fluctuations that will play an important role in the application of quantum information geometry to condensed matter systems. To get a sense of what is meant by this distinction, we introduce the following example, which we refer to as the \emph{classical case}. Imagine a collection of $N$ rotors obeying the laws of classical mechanics, with orientations, $\vec{n}$, and angular momenta, $\hat{L}$, in three dimensions given by, $x\equiv \{\vec{n}_1,...,\vec{n}_N,\hat{L}_1,...,\hat{L}_N\} \in \Omega$, where $\Omega$ is the classical phase space.
The time dependence of the state can be found by solving the classical equations of motion for a given set of initial conditions, yielding a particular trajectory $\chi(t)$. Then, any macroscopic property of the system can be described by the function,  $\Lambda\{\chi(t)\}$. In practice, it is intractable in most cases to specify the initial conditions for any macroscopic collection of particles. Even theoretically, solving the equations of motion is only possible for systems with sufficient symmetry. Thus, in both theory and practice, we consider averages of this function over either fixed intervals of time, or ensembles of possible states, which are equivalent as long as the conditions for ergodicity are met~\cite{birkhoff1931proof}. Classical systems are thus modelled by a probability distribution $p(x)$, called a state. Observable properties are characterized by their averages over this state,
\begin{equation}
    \qexp{\Lambda}_p = \int p(x) \Lambda(x) \dd x
\end{equation}
and their variances and covariances,
\begin{subequations}
\begin{align}
    \text{Var}_{p} \{\Lambda\}   &= \qexp{\Lambda^2}_{p} - \qexp{\Lambda}^2_{p} \\
    \text{Cov}_{p} \{\Lambda_1,\Lambda_2\} &= \qexp{\Lambda_1,\Lambda_2}_{p} - \qexp{\Lambda_1}_{p}\qexp{\Lambda_2}_{p} .
\end{align}
\end{subequations}
In the classical case, fluctuations arise out of our ignorance of the precise conditions of the system at any given instant. We call these fluctuations \emph{classical fluctuations}, or sometimes, \emph{incoherent fluctuations}. The distribution for a system of rotors in equilibrium with a bath of inverse temperature $\beta$ is captured by the Gibbs distribution,
\begin{equation}
    p(x) = \frac{e^{-\beta E(x)}}{\mathcal{Z}}
    \label{eq:ClassicalGibbsDistribution}
\end{equation}
where,
\begin{equation}
    \mathcal{Z} = \int e^{-\beta E(x)}\,\dd x.
    \label{eq:ClassicalPartitionFunction}
\end{equation}
Fluctuations that arise due to equilibration with a bath (whether or particles, energy, volume, etc...) are termed \emph{thermal fluctuations}. All thermal fluctuations are classical, though not all classical fluctuations are thermal. 

Classically, a pure state is one that is specified without any degree of uncertainty (so far as our physical theory allows). For any particular solutions of the equations of motion $\chi(t)$, we have the pure state,
\begin{equation}
    p(x;\chi(t)) = \delta(x-\chi(t))
    \label{eq:DefnClassicalPureState}
\end{equation}
An important albeit obvious property of classical pure states is that they are pure on all subsystems. Consider a subsystem $V$ and its complement $\bar{V}$, such that the phase space is decomposed into $\Omega = \Omega_{V}\oplus\Omega_{\bar{V}}$, with points in phase space factorizing as $x = x_{V}\oplus x_{\bar{V}}$ and trajectories as $\chi(t) = \chi_{V}(t)\oplus \chi_{\bar{V}}(t)$. Then the state of the subsystem will be given by,
\begin{equation}
    p_{V}(x_{V};\chi_{V}(t)) = \int_{\Omega_{\bar{V}}} p(x;\chi(t)) \dd x_{\bar{V}}=\delta(x_{V}-\chi_{V}(t)).
    \label{eq:ClassicalPureStateSubsystem}
\end{equation}
As long as our above factorizations are possible, the above state will also be pure. This is what we mean when we say that a classical pure state is pure on its subsystems.

The revelation of quantum mechanics is that, at the time-energy scales of Planck's constant, the properties of even a single particle undergoing the simplest imaginable dynamics are subject to persistent, irrevocable fluctuations. Where classically we defined a point in phase space explicitly in terms of the microscopic degrees of freedom that we might, in principle, know and measure, the quantum state space is defined by a unit norm vector in Hilbert space called the wavefunction, $\ket{\Psi}$ in a Hilbert space $\mathcal{H}$~\footnote{In fact the particular formalism is not so important. We can define quantum systems in classical state space through the Wigner quasi probability distribution function,~\cite{groenewold1946principles,moyal1949quantum}, but, the Kolmogorov axioms must be relaxed to allow for negative probabilities. We can also define quantum states in real spaces as opposed to complex spaces, by imposing additional constraints on the allowed observables~\cite{bengtsson2017geometry}.}. The wavefunction induces a probability distribution on an observable represented a Hermitian operator $\hat{\Lambda}$ via the amplitude of projection onto the eigenbasis of $\hat{\Lambda}$, $p(\lambda) = \qty|\bra{\Psi}\ket{\lambda}|^2$.

Once we grant that the state of a quantum system is given by a vector in a Hilbert space rather than by a complete set of observable values, all the reasoning of the classical case can be applied. Namely, it is practically very difficult to determine precisely what state a quantum many-body system is in at any given moment, and consequently we consider mixtures of quantum systems represented by the density matrix,
\begin{equation}
    \hat{\rho} = \sum_{j} p_{j}\ket{j}\bra{j}
    \label{eq:DefnDensityMatrix}
\end{equation}
where $p_j$ is a classical probability distribution over the pure states labelled by $j$. Such mixtures introduce incoherent fluctuations, that is, fluctuations not caused by the wave-like nature of quantum phenomena. Just as in the classical case, the mixture in Eq.~\ref{eq:DefnDensityMatrix} can be reduced to a pure state, $\ket{\psi}$ by taking $p_j = \delta_{j,\psi}$. In this case, fluctuations arise entirely from quantum mechanical effects, as we shall see. The extent to which a quantum state represented by the density matrix $\hat{\rho}$ is pure is captured by the \emph{purity},
\begin{equation}
    \mathcal{P} = \Tr\{\hat{\rho}^2\}.
\end{equation}
A purity of $1$ is only achieved if $p_j=\delta_{j,\psi}$, corresponding to the density matrix associated to the pure state $\ket{\psi}$. An important example of a quantum mixed state is that of a system governed by a Hamiltonian $\hat{H}$ in thermal equilibrium with a bath of inverse temperature $\beta$, which we refer to as a Gibbs state,
\begin{equation}
    \hat{\rho}_{\text{Gibbs}} = \frac{\exp{-\beta\hat{H}}}{\mathcal{Z}},
    \label{eq:DefnQuantumGibbsState}
\end{equation}
where, 
\begin{equation}
    \mathcal{Z} = \Tr\{\exp{-\beta\hat{H}}\}
\end{equation}
is the partition function. 

This formal transition, from points in classical phase space to vectors in Hilbert space, facilitates the introduction of two new physical principles. The first is the principle of superposition. Whereas the pure states on the right-hand side of Eq.~\ref{eq:DefnDensityMatrix} are combined incoherently, quantum states can also combine coherently in superposition,
\begin{equation}
    \ket{\psi} = \sum_j c_j \ket{j}
    \label{eq:DefnPureState}
\end{equation}
where $c_j = \bra{j}\ket{\psi}$ are complex numbers such that $\sum_j \qty|c_j|^2 = 1$. Unlike the mixture of states in Eq.~\ref{eq:DefnDensityMatrix}, the superposition of states in Eq.~\ref{eq:DefnPureState} preserves the wave-like nature of quantum effects, allowing for the resulting fluctuations to become correlated in ways not possible when considering incoherent superpositions. One way to understand the difference is that the mixture decreases the purity, whereas super position does not. It is the principle of superposition that gives rise to entanglement. The second is the principle of uncertainty. That is the fact that for pairs of observables, say $\hat{\Lambda}_1$ and $\hat{\Lambda}_2$, have a minimum amount of simultaneous fluctuation,
\begin{equation}
    \text{Var}_{\hat{\rho}} \{\hat{\Lambda}_1\}\text{Var}_{\hat{\rho}}\{\hat{\Lambda}_2\} \geq \frac{1}{4}\qty|\left\langle\comm{\hat{\Lambda}_1}{\hat{\Lambda}_2}\right\rangle_{\hat{\rho}} |^2
\end{equation}
Hence, if $\hat{\Lambda}_1$ and $\hat{\Lambda}_2$ do not commute, $\comm{\hat{\Lambda}_1}{\hat{\Lambda}_2}\neq = 0$, this establishes limits
on $\text{Var}_{\hat{\rho}} \{\hat{\Lambda}_1\}$ and $\text{Var}_{\hat{\rho}}\{\hat{\Lambda}_2\}$.
The fluctuations that arise from these two principles are collectively termed \emph{quantum fluctuations}.

An important implication of the principle of superposition is the emergence of entanglement. In direct analogy with the discussion surrounding Eq.~\ref{eq:ClassicalPureStateSubsystem}, we may consider the state of a many-body pure state on one of its subsystems, $V$. In direct analogy with Eq.~\ref{eq:ClassicalPureStateSubsystem} we have,
\begin{equation}
    \hat{\rho}^{\psi}_V = \Tr_{\bar{V}} \{\hat{\rho}^{\psi}\}.
\end{equation}
Unlike in the case of classical pure states, the state $\hat{\rho}^{\psi}_V$ will only have unit purity if the state $\ket{\psi}$ admits the decomposition
$\ket{\psi} = \ket{\phi_{V}}\ket{\phi_{\bar{V}}}$. If $\hat{\rho}^{\psi}$ is not pure, we say that the subsystems $V$ and $\bar{V}$ are entangled in the state $\ket{\psi}$. 

Entanglement comes down to the \emph{producibility} of a state, that is, the extent to which
it can be factored into subsystems. Since the study of entanglement is a major application
of information geometry, we take a moment to formalize this idea. Let $\ket{\Psi}$ be an $N$ body 
state. Now consider a factorization of this state into $M$ subsystems,
\begin{equation}
    \ket{\Psi} = \bigotimes_{l = 1}^M \ket{\psi_l}
\end{equation}
where each subsystem has $N_{l}$ particles with $\sum_{l}N_{l} = N$. Now let $m$ be an integer such that $m\geq N_{l}$ for all $\ell$. We then say that $\ket{\Psi}$ is $m$-producible. 
If a state is $m$-producible but no $(m-1)$-producible, we say that $\ket{\Psi}$ is 
$m$-partite entangled. 

\begin{figure}
    \centering,
    \begin{subfigure}[b]{0.48\textwidth}
         \centering
         \includegraphics[width=\textwidth]{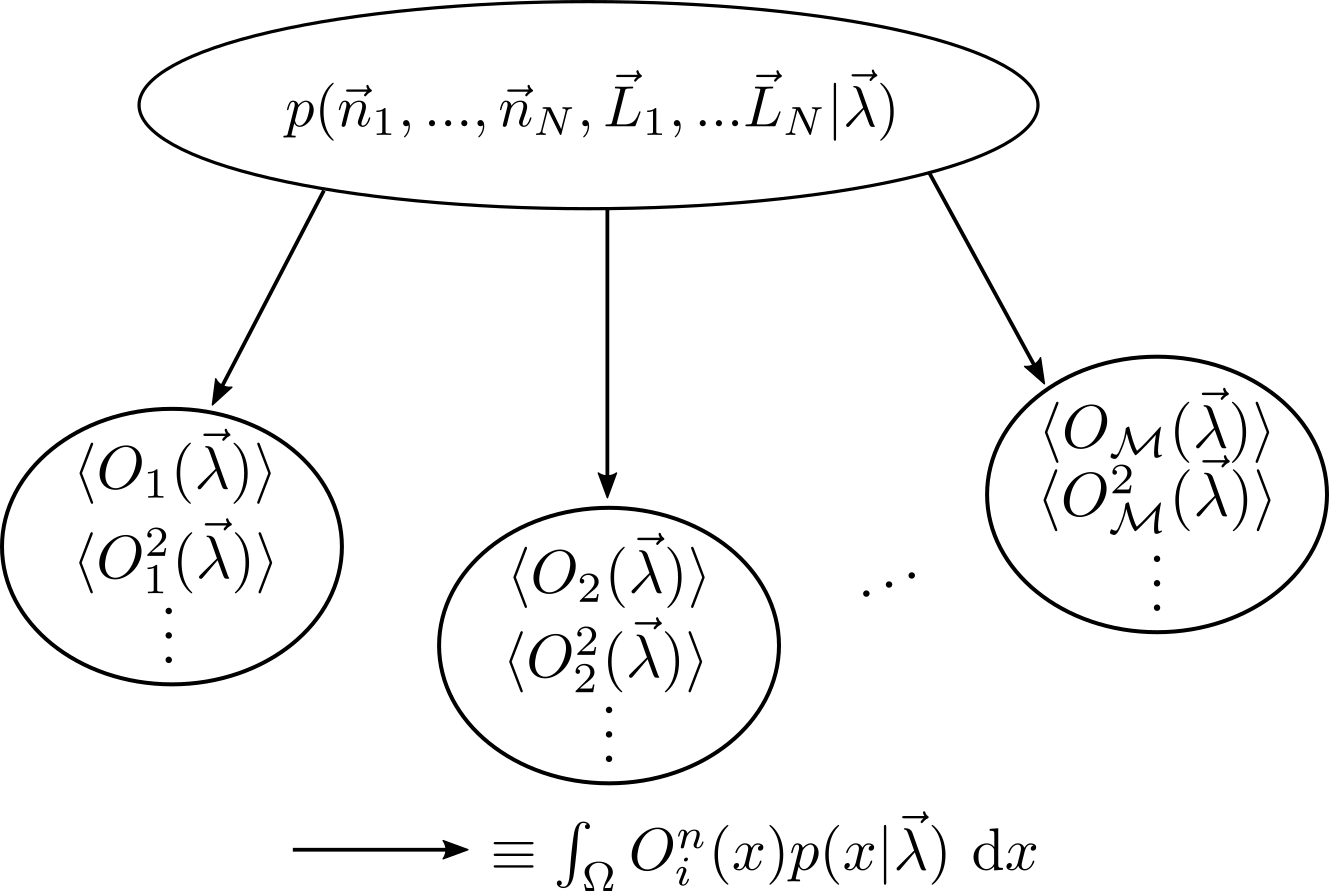}
         \caption{Statistical structure of classical mechanics, with all observable subordinated to a single classical probability distribution.}
        \label{fig:MapOfFluctuations}
     \end{subfigure}
     \hfill
    \begin{subfigure}[b]{0.48\textwidth}
         \centering
         \includegraphics[width=\textwidth]{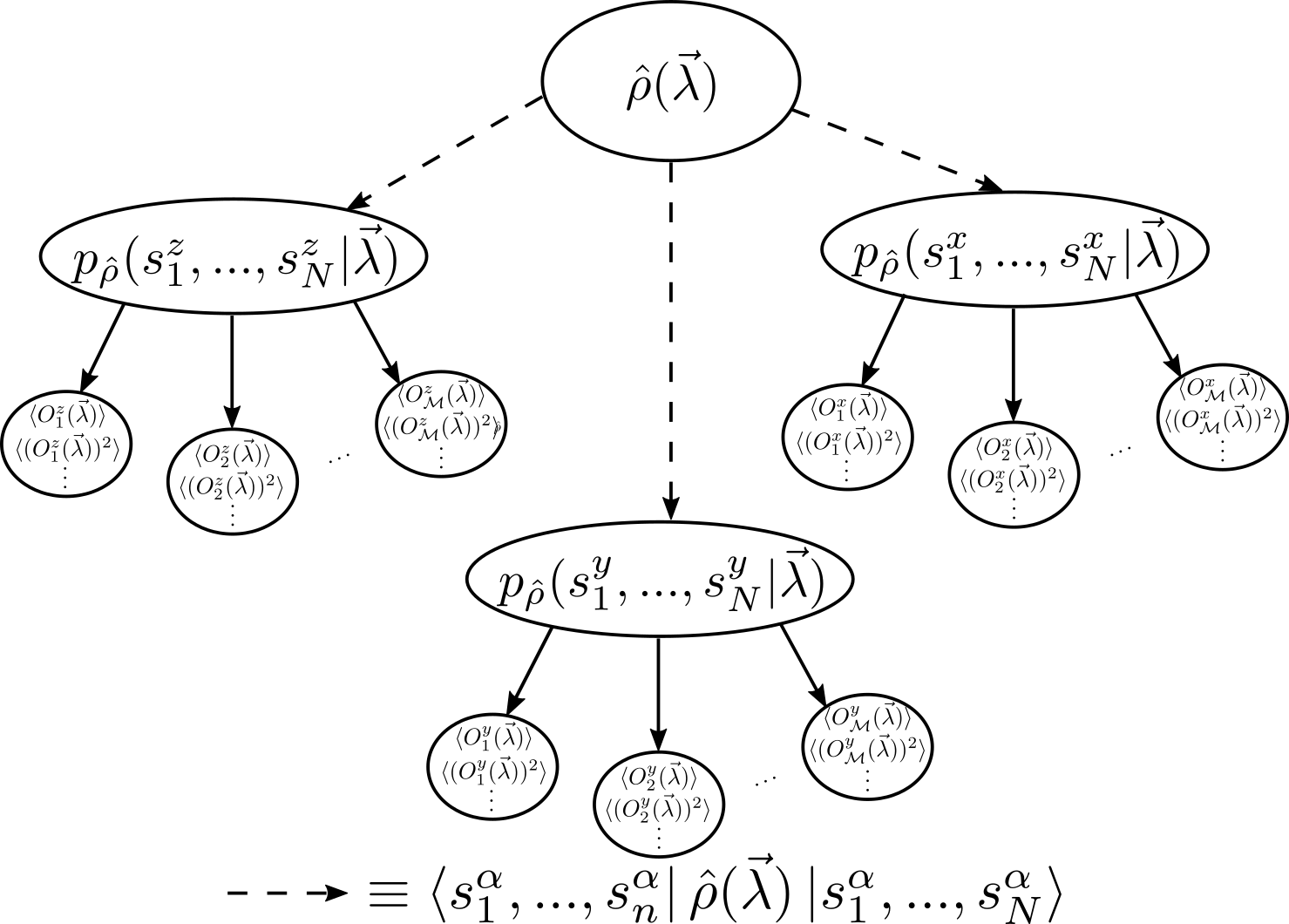}
         \caption{Statistical structure of quantum mechanics, showing the subordination of different sets of observables to different distributions generated by the state $\hat{\rho}$ }
         \label{fig:MapOfQuantumFluctuations}
     \end{subfigure}
\end{figure}

Thus, we have two kinds of fluctuations that coexist in many-body systems. The classical (thermal) fluctuations caused by limitations in our precise knowledge of the state of the system at a given instant, and the quantum fluctuations that seem to be intrinsic to the definition of the states themselves. These quantum fluctuations can give rise to correlations beyond classical physics, of which entanglement is a particular example. The Relationship between the two is depicted schematically in Fig.~\ref{fig:MapOfFluctuations} and Fig.~\ref{fig:MapOfQuantumFluctuations}. In classical statistical mechanics, a single underlying probability distribution governs the outcome of all possible observables by assigning a probability to the classical pure states. In the classical case discussed above this means assigning a probability to each set of orientations and angular momenta. In the quantum case, however, each pure state gives rise to a probability distribution for each non-commuting set of observables. We will return to this fact in Sec. ~\ref{subsubsec:QuantumFisherInformation} when we generalize classical Fisher information to the quantum case. 
Using information theory, we can understand these fluctuations, both classical and quantum, as the notion of distance that induces the aforementioned geometrical structure, and in turn, reveal the presence of quantum correlations. 

The guide is laid out as follows. In the next section, we introduce the concept of quantum information geometry, by beginning with the classical case and then progressing to the quantum case. In section three, we go over key results in several different fields. Finally, we offer some discussion of the overall state of the field and suggest possible new lines of research. In addition, for readers interested in a more thorough discussion of many aspects of quantum information geometry, we note here some textbook length works that we recommend and cite throughout ~\cite{Amari2000,Amari2016,bengtsson2017geometry}.

\section{Introduction to Information Geometry}\label{sec:TheoryOfInformationGeometry}

The study of information geometry, in both the quantum and classical contexts, has a rich mathematical history that predates quantum mechanics itself, with the work of Fubini~\cite{fubini1904sulle} and Study~\cite{study1905kurzeste}, who first studied the Riemannian structure of complex projective spaces. Soon after this, the notion of Fisher information~\cite{fisher1925theory} emerged, which quantifies the precision with which a parameter can be estimated based on measurements sampled from a distribution which depends on the parameter (e.g. we might try to estimate the mean of a Gaussian distribution based on the sample average). The idea of Fisher information explicitly developed into a geometrical idea with the work of Rao~\cite{rao1945information}, and hence, the classical notion of information geometry was born. Some time later, mathematical physicists made the final connection, generalizing the classical Fisher information to the quantum context and, in the process, rediscovering the Fubini-Study metric~\cite{provost1980riemannian,wootters1981statistical}. For a thorough review of the history of information geometry, one can refer to the bibliographic notes in~\cite{brody2008vapourliquid}.

Before introducing the notion of Fisher information, which will lead ultimately to our geometrical picture of the space of probability distributions, we review the more family concept of Shannon information. 

\subsection{Classical Information Geometry}\label{subsec:ClassicalInformationGeometry}

\subsubsection{Shannon Information}\label{subsubsec:ShannonInformation}

In developing our notions of information, it helps to consider a particular problem.
We imagine a meteorological laboratory with a number of field stations in various different locations. Each hour, the stations transmit the current weather conditions. The transmissions occur in binary, with each possible forecast in the set $\{x_i\}_{i=1}^D$ having encodings with lengths $\{L_i\}_{i=1}^D$. For example,  $x_1$, which might correspond to clear skies, could be encoded as $001$ (with $L_1=3$), while $x_2$, which might correspond to overcast skies, could be encoded as $0100$ (with $L_2=4$), and so on.  
 When determining an encoding of forecasts into binary, it is important that we ensure that no forecast's encoding is contained in the prefix of another forecast's encoding, lest we confuse our receiver (for example we don't want to use $1$ as an encoding for clear skies if $10$ is an encoding for cloudy skies). Codes with this property are called prefix codes, and they are always possible provided that the lengths of the encodings, $\{L_1,...,L_D\}$, satisfy the inequality~\cite{kraft1949device},
\begin{equation}
    \sum_{i=1}^D 2^{-L_i} \leq 1
    \label{eq:KraftInequality}
\end{equation}
This inequality can be generalized to alphabets of any size by replacing $2$ with the length of the alphabet. 
In the interest of efficiency, we should aim to assign the shortest encoding to the most likely forecast, so that on average we transmit as few bits as possible. If the probabilities for each forecast are given by the set $\{p_i\}_{i=1}^D$, then we want to minimize the expected length of a transmission,
\begin{equation}
    \qexp{L} = \sum_{i=1}^D p_iL_i
    \label{eq:AveLen}
\end{equation}
subject to the constraint given by Eq.~\ref{eq:KraftInequality}. In doing so, we find that the optimal encoding length of a forecast occurring with probability $p_i$ is, 
\begin{equation}
    \ell_i=-\log_2(p_i).
    \label{eq:LogLikelihood}
\end{equation}
In general we might imagine that our probabilities, $p_i = p(x_i|\vec{\lambda})$ depend on some other parameters, $\vec{\lambda}$ (such as temperature). In this case, Eq.~\ref{eq:LogLikelihood} is called the \emph{log-likelihood}. It can be read as the logarithm of the likelihood that $\vec{\lambda}$ has a particular value given the observed result $x_i$. We will encounter it again shortly. 
From Eq.~\ref{eq:AveLen} we can see that the optimal average message length is bounded from below by,
\begin{equation}
    \qexp{L}\geq \sum_{i=1}^D p_i\ell_i.
\end{equation}
This lower bound is our first measure of information,
\begin{equation}
    S = -\sum_i p_i\log_2(p_i)
    \label{eq:ShannonInformation}
\end{equation}
and, as it was introduced by Claude Shannon~\cite{shannon1948mathematical} a little less than a year before Kraft introduced the inequality in Eq.~\ref{eq:KraftInequality}~\cite{kraft1949device}, we call it the \emph{Shannon Information}.
To gain an understanding of its properties, let's first consider the trivial case where there is a single possible forecast with probability one hundred percent, i.e. $p_i = \delta_{1i}$ where $\delta_{1i}$ is the Kronecker delta function. Then the shortest possible encoding has average length zero. After all, what would be the point in even transmitting? 
Assuming now a uniform probability distribution on $s_i$, we see that the average length of the encoding will grow monotonically with the number of forecasts, $N$. This property is called monotonicity, and it is a natural property to demand from a measure of information. 

The Shannon Information can tell us much more, however, than the average length of an optimal encoding. Considering the case of two possible forecasts, it is clear that the minimal encoding will be one bit (say dot for $x_1$ and dash $x_2$), regardless of the relative values of $p_1$ and $p_2$. If, however, we know that $p_1$ is very nearly 1 and $p_2$ is very close to zero, we can see that the Shannon information tells us something about how much the receiver learns from reading the message. If, at station $P$, the forecast is sunny nine out of every ten days, we learn relatively little when we receive a forecast of sunshine. 
The power of Shannon information is that it captures an intuitive aspect of information: the less probable an event is thought to be, the more information is gained when that event is observed. 

For the mathematically inclined reader, one may wonder to what degree Shannon information is unique. In 1961, Alfred R\'{e}nyi~\cite{renyi1961measures} found an entire family of information measures by imposing a number of conditions (see App.~\ref{app:PropertiesOfInformation}) that are considered sensible for a measure of information. These so-called R\'{e}nyi entropies contain the Shannon information as a special case. The Shannon information is singled out by its property of recursion (its appearance in the prefix coding problem above). 

Physicists will recognize that the Shannon information is a generalization of the concept of entropy. There is no particular reason why we should have to measure information in bits, and we can just as well take the logarithm in Eq.~\ref{eq:ShannonInformation} to have base $e$, (in which case we say we are measuring information in \emph{nats}). Given a thermodynamic ensemble of $W$ possible microstates, all taken to be equally likely with $p_i =\frac{1}{W}$, we see that the Shannon information is simply the Boltzmann entropy,
\begin{equation}
    S = k\ln(W).
    \label{eq:BoltzmannEntropy}
\end{equation}
For this reason we tend to use the words entropy and information interchangeably. Since changing the base of the logarithm simply amount to a redefinition of the units of information, we will do so frequently in the following text as a matter of convenience.

\subsubsection{Fisher Information}\label{subsubsec:FisherInformation}

At our meteorological laboratory, we are interested in producing models that determine the distribution of forecasts from our field stations. To this end, we make a statistical model $P(\vec{\lambda})= \{p_i(\lambda)\}$ where $\vec{\lambda}\in \mathbb{R}^d$ is a parameterization of possible distributions. As we receive data from the field stations, we want to determine as accurately as possible the values of $\vec{\lambda}$ that fit the data. For simplicity, let's consider the case of only one parameter. Let $\lambda$ represent a parameter such as a temperature, and let $\tilde{\lambda}$ be an estimator for this parameter (such as the sample mean of some number of temperature measurements). The function $p(x|\lambda)$ gives the likelihood of the temperature $x$ given the value of the mean temperature in the model $\lambda$. We assume our estimator is unbiased, which means that $\qexp{\tilde{\lambda}} =\lambda$ (this condition can be relaxed without substantially changing the discussion),
\begin{equation}
    0 = \int \dd x \,(\tilde{\lambda}(x) - \lambda)p(x|\lambda).
\end{equation}
Differentiating both sides with respect to $\lambda$, we can write,
\begin{equation}
    0 = \int\dd x \,\,(\tilde{\lambda}(x) - \lambda)\partial_{\lambda}p(x|\lambda) - \int \dd x \, \, p(x|\lambda) .
\end{equation}
Notice that we assume that the estimator itself does not depend on $\lambda$. This makes sense since the estimator is the processing that we do to our data to estimate $\lambda$ and if it depended on $\lambda$ it would be ill-defined. 
Performing the integral in the second term and rearranging we arrive at,
\begin{equation}
    \int\dd x \,\,(\tilde{\lambda}(x)-\lambda) p(x|\lambda)\left(\frac{1}{p(x|\lambda)}\partial_\lambda p(x|\lambda)\right) = 1.
    \label{eq:IntermediateResult}
\end{equation}
Now we notice that the left-hand side of the above equation is just the covariance of the estimator with the logarithmic derivative of the probability. Squaring both sides of the above equation and applying the Cauchy-Schwarz inequality, yields
\begin{equation}
    1 \leq \text{Var}_P\{\tilde{\lambda}\} \text{Var}_P\{\partial_{\lambda}\ln(p(x|\lambda))\}.
\end{equation}
Notice here that we are assuming that the expected value of the derivative of the log-likelihood is zero. This is straightforward to show,
\begin{equation}
    \qexp{\partial_{\lambda}\ln(p(x|\lambda))}=\int p(x|\lambda)\left(\partial_{\lambda}\ln(p(x|\lambda))\right) \dd x
    = \partial_{\lambda}\int p(x|\lambda) \dd x = 0
    \label{eq:LogLiklihoodZero}
\end{equation}
The variance of $\tilde{\lambda}$ is thus bounded from below as,
\begin{equation}
    \text{Var}_{P}(\tilde{\lambda}) \geq \frac{1}{\text{Cov}_{P}
    \left\{\tilde{\lambda},\partial_{\lambda}\ln(p(x|\lambda))\right\}}.    
\end{equation}
Reintroducing the multidimensional nature of the problem and the definition of the Fisher information, we arrive at the Cram\'{e}r-Rao bound,
\begin{equation}
    \text{Cov}_P\{\tilde{\lambda}_{\mu},\tilde{\lambda}_{\nu}\}
    \geq  
    \frac{1}{4F_{\mu\nu}}
    \label{eq:CramerRao}
\end{equation}
where, 
\begin{align}
    F_{\mu\nu} &= \frac{1}{4}\int_{\Omega}\dd x \frac{1}{p(x|\vec{\lambda})}\pdv{p(x|\vec{\lambda})}{\lambda_a}\pdv{p(x|\vec{\lambda})}{\lambda_b}.
    \label{eq:DefnFisherInformation}
\end{align}
is the definition of the \emph{Fisher information}~\cite{fisher1922mathematical}. We've added the factor of $\frac{1}{4}$ for reasons that will become apparent in a moment. The inequality in Eq.~\ref{eq:CramerRao} is known as the Cram\'{e}r-Rao bound~\cite{rao1945information}. It quantifies how much information can be learned about the parameters $\vec{\lambda}$ by sampling from $p(x|\vec{\lambda})$. The appearance of the derivative of the log-likelihood implies a connection between the Shannon information, which is the expectation value of the log-likelihood, and the Fisher information which is the covariance of the partial derivatives of the log-likelihood, and we will explore this connection in more detail momentarily. 

\begin{figure}
    \centering
    \includegraphics[width=0.5\textwidth]{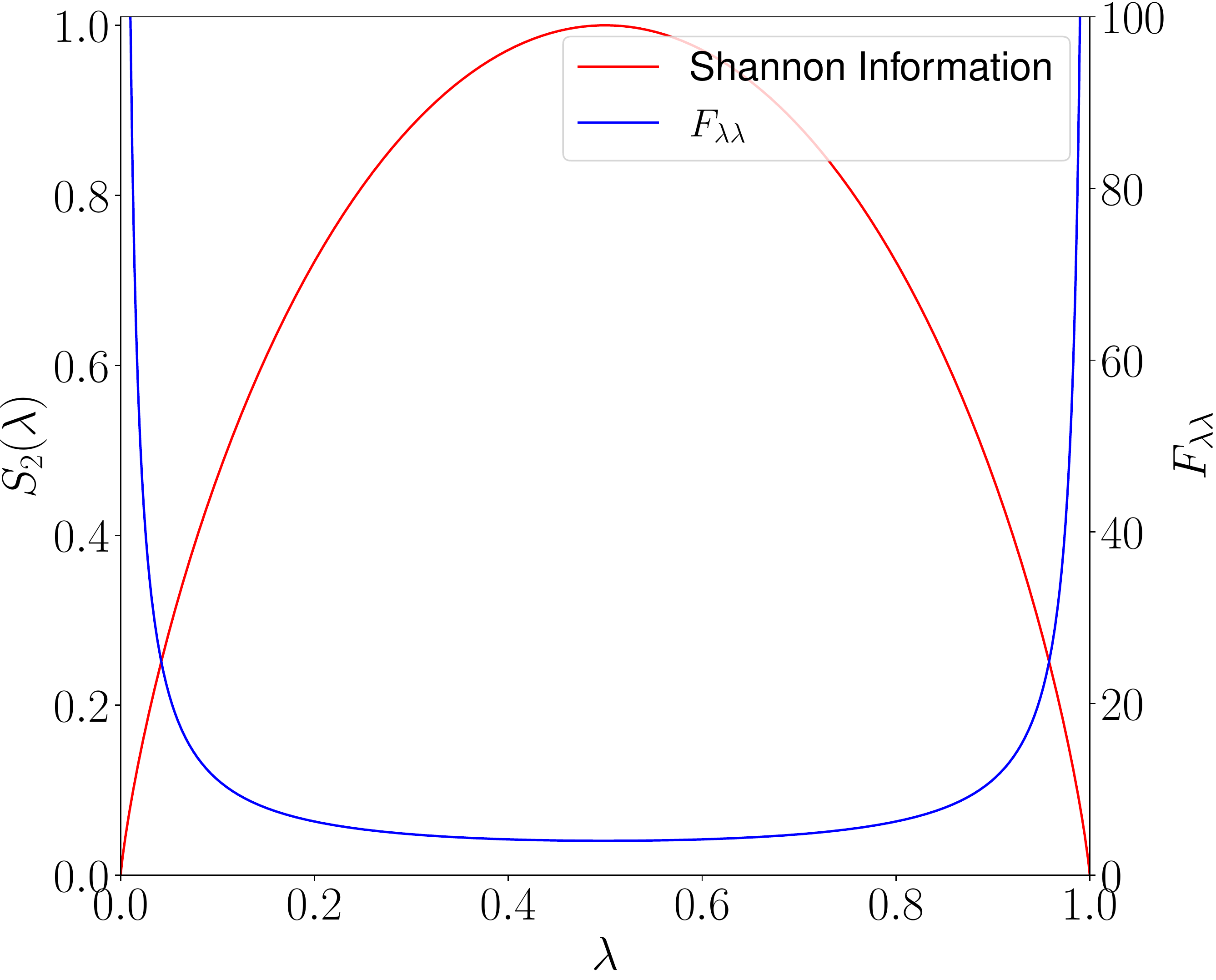}
    \caption[Classical Information of a Coin]{Shannon information (red) in bits and Fisher information (blue) for a weather forecast with $p(\text{Sun})=\lambda$. The location with $\lambda=0.5$ is simultaneously the least predictable (high Shannon information), and the hardest to distinguish from its neighbors (low Fisher information).}
    \label{fig:coinInfo}
\end{figure}

We can begin to get a sense of the meaning of the Fisher information by considering the case of a weather station with only two forecasts. Let $\lambda$ be the probability of sun and $\lambda-1$ be the probability of rain. We can see that when $\lambda=0.5$, the Shannon information -- the information we learn by reading the forecast -- is maximal, whereas the Fisher information -- our ability to estimate $\lambda$ -- is minimal. As $\lambda$ approaches $1$ or $0$, the Shannon information decreases while the Fisher information grows. This behavior is shown in Fig.~\ref{fig:coinInfo} where the Shannon information is given by,
\begin{equation}
    S = -\lambda\log_2(\lambda) - (1-\lambda)\log_2(1-\lambda),
    \label{eq:ShannonInformationCoin}
\end{equation}
and the Fisher information for each measurement is,
\begin{equation}
    F_{\lambda\lambda} = \frac{1}{4} \frac{1}{\lambda(1-\lambda)}.
    \label{eq:FisherInformationCoin}
\end{equation}

\subsubsection{From Information to Geometry}\label{subsubsec:FromInformationToGeometry}

To better understand the connection between Shannon information, Fisher information and geometry, we consider the above scenario, that of estimating the value of the parameter $\lambda$, from a slightly different perspective. 
Considering again the model $p(x|\lambda)$, we can ask how distinguishable $P=p(x|\lambda)$ is from $Q = p(x|\lambda')$.  The degree of distinguishability can be quantified in terms of the informational difference between $P$ and $Q$ by introducing the \emph{relative entropy}~\cite{kullback1951information},
\begin{equation}
    S(P|Q) = \sum_{j=1}^D p_j \log_2\left(\frac{p_j}{q_j}\right)
    \label{eq:DefnKBDivergence}
\end{equation}
If $P=Q$ then the relative entropy is zero, and the distributions are indistinguishable. On the other hand, we can imagine that $P$ corresponds to a place that is always sunny, while $Q$ corresponds to a place that is always rainy. Letting $x_1$ be sunshine and $x_2$ be rain. Then $p_i = \delta_{1i}$ and $q_i=\delta_{2i}$, and we can see that the relative entropy is infinite. 

Let us imagine now that $P$ and $Q$ correspond to the information arriving from different field stations. What are the chances that we might mistake one for the other?
The relative entropy can be used to bound the probability, $E$, that we would obtain a sequence of reports corresponding to the  expected sequences for the distribution, $P$, when sampling $\mathcal{N}$ times from the distribution $Q$,
\begin{equation}
    E \leq (\mathcal{N}+1)^D 2^{-\mathcal{N} S(P|Q)}
    \label{eq:DefnSanovsTheorem}
\end{equation}
This result is called Sanov's theorem~\cite{sanov1958probability}.
In our above example, there is no chance, even on a single transmission, of mistaking the forecast for a town that always rains for a town that is always sunny.

Still considering a situation where the only two possible forecasts are sun or rain, we consider a place $P$ which has, on any given day, a fifty percent chance of either, and compare it to a place $Q$ where the probability of sun is $\lambda$. If we take the limit $\lambda\rightarrow 1$, then the relative entropy $S(P|Q) \rightarrow \infty $. The asymmetry of the relative entropy is apparent here, as $S(Q|P) = 1$. We interpret this to mean that there is zero chance that the typical sequence of $N$ reports from $P$ would be recovered by $Q$ if $Q$ is always sunny. By contrast, there is a chance, which decays with $2^\mathcal{N}$, that the typical sequence of $Q$ will be produced by $P$, albeit one that is vanishingly small. 

This notion of distinguishability is at the heart of information geometry. Intuitively, the more easily distinguishable two distributions, the further they ought to be. This geometry is interesting because it arises naturally in the course of scientific investigation. The example of weather modelling is just one example, and will see several more in the quantum context. But first, we must explain exactly how distinguishability can be used as a metric.

The relative entropy is, asymmetric, and thus unsuitable as a metric in and of itself. However, we can do slightly better if we consider the case where $P$ is varied only infinitesimally. Concretely we define the distributions, $P=\{p_i\}$ and $P+\dd P= \{p_i+\dd p_i\}$, and expand the relative entropy in powers of $\dd p_i$,
\begin{equation}
    S(P|P+\dd P) = \sum_{i=1}^D p_i \ln\left(\frac{p_i}{p_i+\dd p_i}\right) \approx \frac{1}{2}\sum_{i=1}^D \frac{\dd p_i \dd p_i}{p_i}.
\end{equation}
Here we've changed to measuring our relative entropy in nats for convenience. 
This infinitesimal expansion is symmetric, with the asymmetry of the relative entropy entering only at higher order. 
If we imagine that this infinitesimal expansion is a distance, then we can infer the line element,
\begin{equation}
    \dd s^2 = \sum_{i,j=1}^D F_{ij} \dd p_i \dd p_j = \frac{1}{4}\sum_{i,j=1}^D \frac{\dd p_i\dd p_i}{p_i}
\end{equation}
with the \emph{Fisher-Rao metric},
\begin{equation}
    F_{ij} = \frac{1}{4}\frac{\delta_{ij}}{p_i}.
    \label{eq:FisherMetric}
\end{equation} 
We can normalize the metric in any way we like, but the choice of $\frac{1}{4}$ is particularly convenient. To see why, let's make a change of variables $u_i=\sqrt{p_i}$. Now we find, that the line element is,
\begin{equation}
    \dd s^2 = \sum_{i,j=1}^D \dd u_i \dd u_i
\end{equation}
subject to the constraint,
\begin{equation}
    1 = \sum_{i=1}^D u_i^2
\end{equation}
coming from the normalization of the probability distribution. We recognize immediately the equation for an $N$-sphere of unit radius. The constraint that $0\leq u_i\leq 1$ 
confines us to the first octant of this sphere.
Each point on the octant corresponds to a probability distribution, and the geodesic distance between two distributions in terms of the probabilities is simply,
\begin{equation}
    D_{\text{B}}(P,Q) = \cos^{-1}\left(\sum_{i=1}^D \sqrt{p_iq_i}\right)
    \label{eq:BhattacharyyaDistance}
\end{equation}
where we recognize the argument of the inverse cosine as the standard dot product in terms the $u_j$ variables. 
We call this dot product the \emph{classical fidelity} and the corresponding distance the \emph{Bhattacharyya distance}~\cite{bhattacharyya1946measure}. 
 From Eq.~\ref{eq:BhattacharyyaDistance} we can see that pure states are always separated by $\frac{\pi}{2}$. 
 Geometrically, the fact that all classical mixtures of pure points are unique means that the space of classical probability distributions is a simplex\footnote{A simplex is a convex set wherein which each mixture of pure points is unique}.
 We make these points because they will no longer be true when we pass from classical to quantum mechanics.

The Fisher-Rao metric is unique in the sense that it is the only Riemannian metric on the space of probability distributions that is also monotone. The meaning of this is as follows. Let $T$ be a stochastic map, then a monotone function is one for which, given any distributions $P$ and $Q$, $D_B(TP,TQ)\leq D_B(P,Q)$. This property corresponds to the introduction of noise over the channels connecting our laboratory to the field stations. As the channels become noisier, all the incoming transmissions should become less distinguishable. If distinguishability is our notion of distance, then under such stochastic mappings, $\lim_{n\rightarrow\infty} D_B(T^n P, T^n, Q)\rightarrow 0$. 

To complete the connection between the Fisher-Rao metric and the Fisher information derived in the previous section, let us re-express the Fisher-Rao metric in terms of the parameters $\vec{\lambda}$. By the chain rule, the metric in Eq.~\ref{eq:FisherMetric} induces a metric on the space of parameterizations as,
\begin{equation}
    F_{\mu\nu} = \frac{1}{4}\sum_{i,j=1}^D \pdv{p_i}{\lambda_{\mu}}\pdv{p_j}{\lambda_{\nu}}F_{ij}.
\end{equation}
Using our definition of the Fisher-Rao metric we arrive at the \emph{classical Fisher information metric} (CFIM), which we give in both discrete and continuous form,
\begin{subequations}
\begin{align}
    F_{\mu\nu} &= \frac{1}{4}\sum_{i=1}^D \frac{1}{p_i(\vec{\lambda})}\pdv{p_i(\vec{\lambda})}{\lambda_{\mu}}\pdv{p_i(\vec{\lambda})}{\lambda_{\nu}}, \\
    F_{\mu\nu} &= \frac{1}{4}\int_{\Omega}\dd x \frac{1}{p(x|\vec{\lambda})}\pdv{p(x|\vec{\lambda})}{\lambda_{\mu}}\pdv{p(x|\vec{\lambda})}{\lambda_{\nu}}.
\end{align}    \label{eq:DefnCFIM}
\end{subequations}
here $\Omega$ is sample space if $x$ is continuous~\cite{bengtsson2017geometry}. 
Up to the normalization, this metric is the Fisher information that appears in the Cram\'{e}r-Rao bound from Eq.~\ref{eq:CramerRao}. From now on we will write $\pdv{\lambda_{\mu}} \equiv \partial_{\mu}$

There are two other ways of expressing the metric above that we will find useful later. The first takes advantage of the log-likelihood introduced in Eq.~\ref{eq:LogLikelihood}. Notice that Eqs.~\ref{eq:DefnCFIM} are, in fact, covariances in the log-likelihood's partial derivatives (recalling Eq.~\ref{eq:LogLiklihoodZero}), 
\begin{equation}
    F_{\mu\nu} = \text{Cov}_P\left\{\partial_{\mu}\ell,\partial_{\nu} \ell\right\}.
    \label{eq:CovarianceFormOfMetric}
\end{equation}
The covariance can be taken as an inner product, and the derivatives of the log-likelihoods as tangent vectors, allowing us to recover a familiar form of the metric from differential geometry. What is the base space? The Shannon information is the mean of the log-likelihoods, and so we can imagine that the log-likelihood of a probability $p_i$ is the information conveyed by that particular event. The derivatives of the log-likelihood are then tangent vectors on this space of information. 

For certain special choices of coordinate system, it becomes possible for us to represent the metric in terms of a potential function. While these coordinate systems may not always be physically relevant, they can always be defined over a given space of probability distributions. For example, we might parametrize the space of probability distributions by creating linear mixtures of some subset of probability distributions (for example, the pure points). This is the case of the \emph{mixture family} of probability distributions,
\begin{equation}
    p(x|\vec{\lambda}) = \sum_{\mu=1}^N \lambda_{\mu}p_{\mu}(x),
    \label{eq:MixtureFamily}
\end{equation}
where $p_{\mu}(x)$ are some probability distributions (possible discrete or continuous, possibly mixed or pure) and the parameters $\lambda_{\mu}$ are chosen such that normalization is respected. The two-forecast distribution discussed above can be considered as a mixture family by taking $p_1 = \{p(\text{Sun}) = 1, p(\text{Rain}) = 0\}$ and $p_2 = \{p(\text{Sun})=0, p(\text{Rain}) = 1\}$.
If one considers the mixture family in Eq.~\ref{eq:MixtureFamily} then the metric can be given by the Hessian of the Shannon information~\cite{bengtsson2017geometry},
\begin{equation}
    F_{\mu\nu} = -\frac{1}{4}\partial_{\mu}\partial_{\nu} S.
    \label{eq:PotentialFormOfMetric}
\end{equation}
This relationship is not true in general, in particular it fails whenever the second derivative of the probability distribution with respect to the parameters is non-zero. 
This equation can be checked for the two forecast example by comparing Eq.~\ref{eq:ShannonInformationCoin} and Eq.~\ref{eq:FisherInformationCoin}.

A similar statement can be made when one instead considers the \emph{exponential family} of distributions, 
\begin{equation}
    p(x|\vec{\lambda}) = q(x) \exp\left\{\sum_{\mu=1}^N \lambda_{\mu}H_{\mu}(x) - \psi(\vec{\lambda})\right\},
\end{equation}
which are especially relevant in thermodynamics. Here, $x$ is some state in the phase space $\Omega$, $q(x)$ gives the distribution at $\vec{\lambda}=0$ and $\psi(\vec{\lambda})$ is the normalization which plays the role of the thermodynamic potential or free energy,
\begin{equation}
    \psi(\vec{\lambda}) = \ln\left(\int_{\Omega} \exp\left\{\sum_{\mu=1}^N \lambda_{\mu}H_{\mu}\right\} q(x)\dd x \right)
\end{equation}
In this case we can treat $\psi(\vec{\lambda})$ as the potential for the metric via~\cite{brody2008vapourliquid},
\begin{equation}
    F_{\mu\nu} = -\frac{1}{4}\partial_{\mu}\partial_{\nu}\psi(\vec{\lambda})
\end{equation}
For a detailed discussion on the metric potential one can see Ref.~\cite{bengtsson2017geometry}. For a discussion of the role of that this relationship plays in classical thermodynamics, one can refer to Ref.~\cite{brody2008vapourliquid}.

\subsection{Quantum Information Geometry}
We now generalize the concepts developed above in a classical setting to the quantum case. 

\subsubsection{von Neumann Entropy}\label{subsubsec:vNeumanEnt}
We first consider the quantum analogue of the Shannon information, which will allow us to ultimately quantify the entanglement between subsystems. 
Starting from Eq.~\ref{eq:ShannonInformation}, we can apply a heuristic mapping from the classical to the quantum, 
\begin{align}
    p(x)\rightarrow \hat{\rho}, \quad\quad \Lambda(x)\rightarrow\hat{\Lambda}, \quad\quad \int (\cdot)\, \dd x \rightarrow \Tr{\cdot}.
    \label{eq:ClassicalToQuantumMapping}
\end{align}
We immediately arrive at the von Neumann entropy,
\begin{equation}
    \mathcal{S} = -\Tr\{\hat{\rho}\ln(\hat{\rho})\}
    \label{eq:vonNeumannEntropy}
\end{equation}
where we measure the information in nats. In doing so, we introduce our convention that the quantum analogue of a classical information measure is written in script font. We can see that the von Neumann entropy will be zero for pure quantum states, just as the Shannon entropy was zero for pure classical states. 

At first glance, the von Neumann entropy doesn't seem particularly useful for detecting entanglement, since it will give zero for all pure states. In fact, this is exactly its utility. If the von Neumann entropy of the subsystem state $\hat{\rho}^{\Psi}_{V}$ is non-zero, then we know that the subsystem $V$ is entangled with its complement, $\bar{V}$ as per our earlier discussion. For this reason, the von Neumann entropy of a subsystem $\mathcal{S}_{V}$ is called the \emph{entanglement entropy}. Useful in this interpretation is the fact that $\mathcal{S}_{V} = \mathcal{S}_{\bar{V}}$, that is, $V$ is entangled with $\bar{V}$ exactly as much as $\bar{V}$ is entangled with $V$. This is true even if $V$ and $\bar{V}$ are not the same size, as long as they form a disjoint and complete partition of the system. 

To see how the entanglement entropy works, we can consider the simple example of the entanglement between the two spins for the state,
\begin{equation}
    \ket{\text{GHZ}_N} = \frac{1}{\sqrt{2}}\left(\ket{\uparrow}^{\otimes N} 
                                               + \ket{\downarrow}^{\otimes N}\right)
    \label{eq:DefnGHZState}
\end{equation}
with $N=2$. In this case, the von Neumann entropy between the subsystems will be $1$ bit. 

The entanglement entropy has seen extensive application in quantum information and many-body physics. A significant triumph is the scaling laws of entanglement entropy with subsystem size, which characterize the central charge of conformal field theories in models where such theories describe the low energy degrees of freedom~\cite{Holzhey1994,Calabrese2004,calabrese2009entanglement}. For many-body systems with local interactions, the entanglement entropy tends to scale with the area of the subsystem under consideration~\cite{eisert2010colloquium}. This fact is crucial to the efficiency of the density matrix renormalization group family of algorithms~\cite{White1992b}. 

Taking Eq.~\ref{eq:vonNeumannEntropy} as the expectation value of the operator $-\ln(\hat{\rho})$, we define the latter as the \emph{entanglement Hamiltonian},
\begin{equation}
    \hat{H}_{\text{ent}} = -\ln(\hat{\rho})
\end{equation}
Phases exhibiting symmetry protected topological order have been shown to exhibit non-trivial degeneracies in the spectrum of the entanglement Hamiltonian (the entanglement spectrum)~\cite{Haldane2008,pollmann2010entanglement}. Another significant aspect of the entanglement entropy is that, for phases with long range topological order in two dimensions and greater, the area law scaling of the entanglement entropy is modified by a negative subleading correction, the \emph{topological entanglement entropy}~\cite{kitaev2006topological}.

These facts have made the entanglement entropy ubiquitous as a theoretical probe of quantum fluctuations and correlations. Experimentally, however, the application of Eq.~\ref{eq:vonNeumannEntropy} is quite limited. In particular, computing the entanglement entropy requires complete knowledge of the density matrix, something which is only possible in highly controlled experimental settings.

\subsubsection{Quantum Fisher Information}\label{subsubsec:QuantumFisherInformation}

When generalizing the concept of Fisher information to the quantum case, we must address the fact that quantum mechanics allows us to choose between sets of measurements that are decoherent with one another. For example, we might orient our Stern-Gerlach devices along any spin axis we like, and in doing so will be sampling from distinct probability distributions corresponding to the different choices of measurement basis. To formalize this, we introduce the concept of the \emph{positive operator valued measure} (POVM). See for instance \cite{holevo2003statistical,holevo2011probabilistic}. Let $\{\ket{\xi}\bra{\xi}\}_{\xi=1}^M$ be a set of positive semi definite operators (hereafter projectors) acting on the Hilbert space satisfying the resolution of the identity,
\begin{equation}
    \mathbb{I} = \sum_{\xi} \ket{\xi}\bra{\xi}
    \label{eq:DefnPOVM}
\end{equation}
In general we do not require that the projectors be orthogonal as they would be if they represented the spectrum of a Hermitian operator. Eq.~\ref{eq:DefnPOVM} represents the most general possible kind of measurement that can be made on a quantum state. For each set of observables that commute with each other, there exists a corresponding POVM which we might label with an additional superscript, $\alpha$. Referring back to Fig.~\ref{fig:MapOfQuantumFluctuations}, each of the probability distribution branching off of the density matrix correspond formally to,
\begin{equation}
    p(\xi^\alpha) = \Tr\{\ket{\xi^\alpha}\bra{\xi^\alpha}\hat{\rho}\}
\end{equation}
When considering the quantum generalization of the Fisher information, we see that we would in fact have a QFI for each possible choice of POVM. What is conventionally called the QFI then corresponds to maximizing the classical Fisher information over all possible choices of POVM~\cite{braunstein1994statistical,braunstein1996generalized}. 
We make this point only briefly because it is relevant in the case of quantum sensing, which we return to later. For now, we move straight to the quantum generalization of the geometrical structure introduce in Sec.~\ref{subsubsec:FromInformationToGeometry}. 

\subsubsection{Quantum Geometric Tensor}

In contrast to classical states, quantum states can vary in two distinct ways.
To see this we consider a general mixed state depending on the parameter $\lambda$,
\begin{equation}
    \hat{\rho}(\lambda) = \sum_{j} p_j(\lambda) \ket{j(\lambda)}\bra{j(\lambda)}.
    \label{eq:ParamDensityMatrix}
\end{equation}
A change in the parameter $\lambda$ might cause a change in the eigenvalues and/or a change in the eigenvectors. The former of these changes induces a distance identical to the kind that we considered in the classical case. Those parameterizations that change the eigenvectors give rise to quantum information geometry. We can study this case in detail by first considering a pure state, $\ket{\Psi(\vec{\lambda})}$.
Now we can expand this state to leading order for a small change in the component of the parameter $\lambda_{\mu}$, 
\begin{equation}
    \ket{\Psi(\lambda_{\mu}+\dd\lambda_{\mu})} = \ket{\Psi(\lambda_{\mu})} + \dd\lambda_{\mu} 
    \ket{\partial_{\mu}\Psi(\lambda)},
\end{equation}
where $\partial_{\mu} \equiv \partial_{\lambda_{\mu}}$.
Intuitively, the distance between states is the size of the change in the state. We need to take care however to ensure that our notion of distance is independent of the global phase, since physical states are defined by rays in the Hilbert space. We thus define the \emph{quantum geometric tensor} (QGT),
\begin{equation}
    \mathcal{Q}_{\mu\nu} = \bra{\partial_{\mu}\Psi} (\mathbb{I}-\ket{\Psi}\bra{\Psi})\ket{\partial_{\mu}\Psi}
    \label{eq:QuantumGeometricTensor}
\end{equation}
where we've subtracted off the projection of the change in the state $\ket{\partial_{\mu}\Psi}$ onto the subspace of the unchanged state defined by the projector $\ket{\Psi}\bra{\Psi}$. 

The object in Eq.~\ref{eq:QuantumGeometricTensor} admits a decomposition into a real symmetric component and an imaginary antisymmetric component,
\begin{equation}
    \mathcal{Q}_{\mu\nu} = \mathcal{F}_{\mu\nu} + i\Omega_{\mu\nu}.
\end{equation}
The real part, $\mathcal{F}_{\mu\nu}$ gives the Riemannian metric on the quantum state space~\cite{fubini1904sulle,study1905kurzeste}, while the imaginary antisymmetric part, $\Omega_{\mu\nu}$ gives the Berry curvature~\cite{berry1984quantal}. 
From this point on, we focus in particular on the real symmetric component, which we term the \emph{quantum Fisher information metric} (QFIM). The reason for this decomposition is that the Hilbert space in which we've formulated our geometry is complex. Consequently, the metric tensor is required to be Hermitian, but not purely real. The imaginary component will not contribute to the line element due to the fact that it is antisymmetric. The state space thus has two parallel geometric structures. On the one hand, the real Riemannian structure, $\mathcal{F}_{\mu\nu}$, and on the other hand the \emph{symplectic} structure determined by $\Omega_{\mu\nu}$

The distance on the space of pure states is a direct analogue of the Bhattacharyya distance in Eq.~\ref{eq:BhattacharyyaDistance}~\cite{wootters1981statistical},
\begin{equation}
    D_{\text{QB}} (\Psi_1,\Psi_2) = \cos^{-1}\left(\qty|\bra{\Psi_1}\ket{\Psi_2}|\right)
    \label{eq:DefnQuantumBhattacharyyaDistance}
\end{equation}
where the argument of the inverse cosine is now the quantum analogue of the fidelity. In Ref.~\cite{jozsa1994fidelity}, the \emph{quantum fidelity} is defined at $|\bra{\Psi_1}\ket{\Psi_1}|^2$ in place of the classical fidelity. The distance between pure states is sometimes called the \emph{quantum angle}. Where classically, pure states are always separated by the maximal distance of $\frac{\pi}{2}$, now it is only orthogonal pure states that are separated by $\frac{\pi}{2}$. Instead of the positive octant of a hypersphere, the space of quantum pure states can be visualized as a Bloch hypersphere, with anti-podal points identified as orthogonal states. 

All physical pure states must be normalized, and so in the pure state limit only unitary parameterizations are relevant. Without loss of generality, we can always express a unitary parameterization via the Schr\"{o}dinger equation,
\begin{equation}
    \partial_{\mu} \ket{\Psi} = -i\hat{\Lambda}_{\mu}\ket{\Psi}
    \label{eq:SchrodingerEquation}
\end{equation}
where $\hat{\Lambda}_{\mu}$ is a Hermitian operator corresponding to a change in the parameter $\lambda_{\mu}$. By substituting Eq.~\ref{eq:SchrodingerEquation} into Eq.~\ref{eq:QuantumGeometricTensor} we see that the QFIM for unitary parameterizations is given by,
\begin{equation}
    \mathcal{F}_{\mu\nu}^{\text{unitary}} = 4\Re[\text{Cov}_{\Psi}\{\hat{\Lambda}_{\mu},\hat{\Lambda}_{\nu}\}]
    \label{eq:DefnPureStateQFIM}
\end{equation}

For mixed states, unitary parameterizations are slightly more subtle. As a first attempt, we may try to take Eq.~\ref{eq:DefnCFIM} and directly apply the mapping in Eq.~\ref{eq:ClassicalToQuantumMapping}. The problem with this is that, in general, $\comm{\hat{\rho}}{\partial_{\mu}\hat{\rho}} \neq 0$. Resolving this ambiguity is, to a degree, a matter of convention.  We proceed along the lines of Braunstein and Caves~\cite{braunstein1994statistical}, who resolve this ambiguity by introducing the superoperator,
\begin{equation}
    \mathcal{L}_{\hat{\rho}}^{-1}\{\hat{O}\} = \frac{1}{2}(\hat{\rho}\hat{O} + \hat{O}\hat{\rho}).
    \label{eq:SuperOperator}
\end{equation}
In the eigenspace of the density matrix, the inverse of this operator is given by,
\begin{equation}
    \mathcal{L}_{\hat{\rho}}(\hat{O}) = \sum_{j,k} \frac{2}{p_j + p_k} O_{jk}\ket{j}\bra{k},
\end{equation}
with the sum running only over terms with $p_j+p_k > 0$.
Notice that the above form is the inverse of Eq.~\ref{eq:SuperOperator} in the sense that $\mathcal{L}_{\hat{\rho}}\{\mathcal{L}^{-1}_{\hat{\rho}}\{\hat{O}\}\}=\hat{O}$.
Applying the inverse to the derivative of the density matrix,
\begin{equation}
    \mathcal{L}_{\hat{\rho},\mu} = \mathcal{L}_{\rho}\left(\pdv{\hat{\rho}}{\lambda_{\mu}}\right)
    \label{eq:DefnSymmetricLogarithmicDerivative}
\end{equation}
gives us a logarithmic derivative wherein the ambiguity in the ordering of the operators is resolved symmetrically, hence we call Eq.~\ref{eq:DefnSymmetricLogarithmicDerivative} the \emph{symmetric logarithmic derivative} (SLD). Now we can write the QFIM for mixed states as,
\begin{equation}
    \mathcal{F}_{\mu\nu} = 4\Re\left[\text{Cov}_{\hat{\rho}}\{\mathcal{L}_{\hat{\rho},\mu},\mathcal{L}_{\hat{\rho},\nu}\}\right].
    \label{eq:DefnQfiMetricSymLogDeriv}
\end{equation}
The above equation can be taken as the quantum analogue of Eq.~\ref{eq:CovarianceFormOfMetric}. It includes the special case where only the eigenvalues of the density matrix are modified, and in that instance reduces to the classical Fisher information matrix (CFIM). Parameterizations that modify the eigenvectors of the density matrix are captured by the von Neumann equation,
\begin{equation}
    \partial_{\mu}\hat{\rho} = i\comm{\hat{\rho}}{\hat{\Lambda}_{\mu}}.
    \label{eq:vonNeumannEquation}
\end{equation}
which described unitary evolutions generated by the Hermitian operator $\hat{\Lambda}_{\mu}$

As we mentioned, the choice of resolving the ambiguity in ordering through the SLD is just one choice. 
Morozova and \v{C}encov~\cite{morozova1989markov} demonstrated that for every operator monotone function, we can define a metric on the space of mixed states. Skipping the rigorous details, we present here the relationship between metrics and operator monotone functions. 
First we introduce the superoperators~\cite{shitara2016determining},
\begin{subequations}
\begin{align}
    R_{\hat{\rho}}(\hat{O}) &= \hat{O}\hat{\rho} \\
    L_{\hat{\rho}}(\hat{O}) &= \hat{\rho}\hat{O} \\
    (\mathcal{L}_{\hat{\rho}}^f)^{-1}        &= f(L_{\hat{\rho}}R^{-1}_{\hat{\rho}})R_{\hat{\rho}}.
\end{align}
\end{subequations}
Here $f$ is the operator monotone function. For the case of the SLD, we can find that, 
\begin{equation}
    f^{\text{SLD}}(x) = \frac{1+x}{2}.
    \label{eq:SLDOperatorMonotoneFunction}
\end{equation}
There are many other possible choices (we list several in Tab.~\ref{tab:QFIMs}). These types of operator monotones were studied extensively in the work of D\'{e}nes Petz, see 
Ref.~\cite{petz1996monotone,petz1996geometries,petz1996riemannian}. 

The QFIM 
is distinguished by the fact that it finds the maximal distance between states, and is, in this sense, unique~\cite{bengtsson2017geometry}.

\begin{table}
\centering
\begin{tabular}{|c | c |  }
\hline
    QFIM  &  monotone function \\
    \hline 
    \hline
    Symmetric logarithmic derivative &  $\frac{1 + x}{2}$ \\
    \hline
    Bogoliubov-Kubo-Mori & $\frac{x-1}{\log(x)}$ \\
    \hline
    right logarithmic derivative & $x$ \\
    \hline
    left logarithmic derivative & $1$ \\
    \hline
    Skew information ($\alpha=1$) & $\frac{(\sqrt{x}+1)^2}{4}$\\
    \hline
\end{tabular}
\caption[QFIM's for different monotone functions]{A partial reproduction of the table in~\cite{shitara2016determining} listing several different forms of the QFIM along with the associated monotone function.}
\label{tab:QFIMs}
\end{table}

The SLD QFIM can be used to define the Bures distance between mixed states in analogy with Eq.~\ref{eq:DefnQuantumBhattacharyyaDistance}~\cite{bures1969extension},
\begin{equation}
    D_{\text{Bures}} (\hat{\rho}_1,\hat{\rho}_2) = \cos^{-1}\left(\sqrt{F(\hat{\rho}_1,\hat{\rho}_2)}\right),
    \label{eq:DefnBuresDistance}
\end{equation}
where 
\begin{equation}
    F(\hat{\rho}_1,\hat{\rho}_2) = \Tr(\sqrt{\sqrt{\hat{\rho}_1}\hat{\rho}_2\sqrt{\hat{\rho}_1}})^2
    \label{eq:DefnUhlmannFidelity}
\end{equation}
is the mixed state generalization of the quantum fidelity, often termed the Uhlmann fidelity~\cite{uhlmann1976transition}. The above expression can be interpreted as the transition amplitude between mixed states, in exactly the same way as the quantum fidelity can be interpreted as the transition amplitude between pure states. 
Taking the SLD QFIM for granted, the Bures distance is the broadest generalization of Riemannian distance on the quantum state space.

Focusing on the symmetric part of Eq.~\ref{eq:QuantumGeometricTensor} it is useful to explicitly decompose the QFIM into a component that arises due to changes in the eigenvectors of the density matrix (the unitary component), and a component that arises due to changes in the eigenvalues (the non-unitary component) (see for example the supplementary material of Ref.~\cite{hauke2016measuring}),
\begin{equation}
    \mathcal{F}_{\mu\nu} = \mathcal{F}_{\mu\nu}^{\text{non-unitary}} + \mathcal{F}_{\mu\nu}^{\text{unitary}},
    \label{eq:UnitaryNonUnitaryDecomp}
\end{equation}
where,
\begin{equation}
    \mathcal{F}^{\text{non-unitary}}_{\mu\nu} = \frac{1}{4}\sum_j \frac{1}{p_j} \partial_\mu p_j \partial_\nu p_j,
    \label{eq:NonUnitaryComponent}
\end{equation}
and 
\begin{equation}
    \mathcal{F}_{\mu\nu}^{\text{unitary}} = 2\sum_{j,k} \frac{(p_j-p_k)^2}{p_j+p_k} [\Lambda_\mu]_{jk}[\Lambda_\nu]_{kj}^*.
    \label{eq:DefnQFIMUnitaryPart}
\end{equation}
Eq.~\ref{eq:NonUnitaryComponent} is the direct analogue of Eqs.~\ref{eq:DefnCFIM}. Eq.~\ref{eq:DefnQFIMUnitaryPart} has no direct classical analogue. It emerges specifically due to the non-commutativity of quantum observables. Often times, the expression in Eq.~\ref{eq:DefnQFIMUnitaryPart} is called the quantum Fisher information, due to this fact. To maintain our symmetry with the notation used in the discussion of von Neumann entropy, we prefer to call the entire metric, containing both unitary and non-unitary components the QFIM, and instead refer to the unitary and non-unitary components as needed. Because the Berry curvature is antisymmetric, there is no such ambiguity. 

\subsubsection{Topological Invariants and Volumes}

The QFIM allows for the construction of two distinct but related topological invariants. We discuss these invariants specifically in the case of a two-dimensional submanifold of states, $\mathcal{M}$, which might be given by the parameters of a Hamiltonian, or by the first Brillouin zone.

The first emerges from the Berry curvature, which, when integrated over a two-dimensional manifold, gives the Chern number of the manifold,
\begin{equation}
    C = \frac{1}{2\pi}\int_{\mathcal{M}} \Omega_{\mu\nu}\dd\lambda_{1}\dd\lambda_{2}.
    \label{eq:ChernNumber}
\end{equation}
The second is the Euler characteristic, which is given by Gauss-Bonnet theorem~\cite{do2016differential},
\begin{equation}
    \chi = \frac{1}{2\pi}\left(\int_{\mathcal{M}} K \,\,\dd A
    +\oint_{\partial\mathcal{M}} k_g \,\,\dd\ell
    \right).
\end{equation}
In the above expression, $K$ is the \emph{Gaussian curvature} and $k_g$ the \emph{geodesic curvature} of the boundary. In order to express these explicitly in terms of the QFIM, we model our notation on Ref.~\cite{kolodrubetz2013classifying}.
Let us write the metric in first fundamental form, by expressing the line element as
\begin{equation}
    \dd s^2 = E\dd\lambda_{1}^2 + 2F \dd\lambda_{1}\dd\lambda_{2} + G\dd\lambda_{2}.
\end{equation}
We denote the determinant of the metric by omitting the Greek indices,
\begin{equation}
    \mathcal{F} = \det{\mathcal{F}_{\mu\nu}} = EG-F^2.
\end{equation}
We also recall the Christoffel symbols,
\begin{equation}
    \tensor{\Gamma}{^k_i_j} = \frac{1}{2}\tensor{\mathcal{F}}{^k^m}
    \left(
        \partial_j\tensor{\mathcal{F}}{_i_m}
    +   \partial_i\tensor{\mathcal{F}}{_j_m}
    -   \partial_m\tensor{\mathcal{F}}{_i_j}
    \right).
\end{equation}
Now we can express the Gaussian curvature and the geodesic curvature, 
\begin{subequations}
    \begin{align}
        K &= \frac{1}{\sqrt{\mathcal{F}}}
        \left( \partial_2\left(\frac{\sqrt{\mathcal{F}}\,\tensor{\Gamma}{^2_1_1}}{E}\right)
        - \partial_1\left(\frac{\sqrt{\mathcal{F}}\,\tensor{\Gamma}{^2_1_2}}{E}\right)
        \right) \\
        k_g &= \sqrt{\mathcal{F}} G^{-\frac{3}{2}}\tensor{\Gamma}{^1_2_2},
    \end{align}
\end{subequations}
along with the line and area elements,
\begin{subequations}
    \begin{align}
        \dd A &=\sqrt{\mathcal{F}}\dd\lambda_1\dd\lambda_2 \\
        \dd \ell &= \sqrt{G}\dd\lambda_2
    \end{align}
\end{subequations}
The metric can of also be used to evaluate the volume of the parameter manifold in $n$ dimensions via,
\begin{equation}
    \text{Vol}_g = \int \sqrt{\det(g)}\dd\lambda_1\wedge\dd \lambda_2\wedge\ldots\wedge\dd\lambda_n
    \label{eq:Volume}
\end{equation}
In Sec.~\ref{subsubsec:ChernNumberAndTopo} and Sec.~\ref{subsubsec:StateSpaceCurvature} we offer further discussion of the analysis of these topological invariants in several systems.

\section{Information Geometry in Practice}\label{sec:PhyiscsOfInformationGeometry}

So far, we have laid out the geometrical structure of the quantum state space. In this section, 
we try to understand what this structure means from a physical perspective, with a particular
focus on the applications of this structure to many-body systems. There are two
main topics to be addressed here. The first is understanding what, in physical terms, the
quantum information geometry is actually describing. A hint is already given by the fact
that the metric is the real part of the covariance. The second is understanding several fundamental choices of parameter manifold, and the contexts in which these emerge. 

For the reader's convenience, we identify three main categories of manifolds that appear in the literature. The first is the Brillouin zone in a band system parametrized by the momenta $(k_x,k_y)$ with $k_{\alpha}\in[0,2\pi)$. The non-trivial topology of the band structure, which leads to the classification of topological phases of matter, emerges from the symplectic component of the QGT: the Berry curvature. Several results in this direction have found interested geometrical properties associated with this geometry~\cite{sarkar2022free,kolodrubetz2017geometry,ma2013euler,ozawa2018extracting,mera2021kahler,mcclarty2021topological,gianfrate2020measurement,tan2019experimental,kemp2021nested}.

In the second case are manifolds of ground state parametrized by fields and couplings in a parent Hamiltonian (see, for example, Refs.~\cite{thesberg2011general,tzeng2008scaling,albuquerque2010quantum,gu2008fidelity,gu2009scaling,gu2010fidelity,zanardi2006ground,zanardi2007bures,zanardi2007information,you2007fidelity,yang2008fidelity,garnerone2009fidelity,wang2015fidelity,cozzini2007quantum,langari2013ground,Schwandt2009,mera2018dynamical}). This kind of geometry emerges especially in the context of the Fidelity susceptibility, and we treat in some details the structure of the QGT in Sec.~\ref{subsubsec:FidelitySusceptibility}. Finally, in the context first of metrology and increasingly in many body physics, are manifolds generated by the application of some unitary operator $\hat{U} = \exp{-i\lambda\hat{\Lambda}}$ to the state $\ket{\psi}$ with $\hat{\Lambda}$ a bounded Hermitian operator with spectrum.

In the case of metrology, $\hat{\Lambda}$ might play the role of the coupling between the probe state and the system being studied, with $\lambda$ the parameter to be estimated. In the context of many body spin systems, $\hat{\Lambda}$ is usually a sum over operators defined on each degree of freedom in the system, such as the collective spin operator,
\begin{equation}
    \hat{\Lambda} = \sum_j S_j^{\alpha}
    \label{eq:LocalOp}
\end{equation}
 along some particular axis $\alpha\in\{x,y,z\}$ (see for example Ref.~\cite{lambert2020revealing,lambert2019estimates,lambert2022state,hauke2016measuring,yin2019quantum,laurell2021quantifying,scheie2021witnessing,li2013spin,ma2009fisher,liu2013quantum,toth2012multipartite,hyllus2012fisher}). In this case the components of the quantum metric correspond to the integrated dynamical response of the system. Operators of the form given in Eq.~\ref{eq:LocalOp} are called local, because the summand depends on a bounded number of sites even in the thermodynamic limit. Along this line of reasoning, Ref.~\cite{pezze2017multipartite} found super-extensive scaling of the QFI in the topologically non-trivial Kitaev wire by choosing a non-local operators,
 \begin{equation}
     \hat{\Lambda}^{(\alpha)} = \sum_j (-i)^{\delta_{\alpha,x}}
     \left(\hat{a}_j^{\dagger} e^{i\pi\sum_{i=1}^{j-1}\hat{n}_l} + (-1)^{\delta_{\alpha,y}} e^{-i\pi\sum_{i=1}^{j-1}\hat{n}}\hat{a}_j\right)
 \end{equation}
where $\hat{a}_j$ is the fermionic annihilation operator and $\hat{n}$ is the number operator. We discuss the relationship between these three branches of research more in the following sections. 
 
\subsubsection{Relationship to Linear Response}\label{subsec:relationToLinearResponse}

Considering the fact that the QFIM quantifies the degree of distinguishability along the path generated by the hermitian operator $\hat{\Lambda}_\mu$, it is not surprising that the unitary component of QFI has a convenient representation in terms of the dissipative part of the linear response~\cite{hauke2016measuring}, 
\begin{equation}
    \mathcal{F}_{\mu\nu}^{\text{unitary}} = \frac{4}{\pi}\int_0^{\infty} \dd\omega\tanh\left(\frac{\omega}{2T}\right)\chi''_{\mu\nu}(\omega).
    \label{eq:QfiLinearResponse}
\end{equation}
where,
\begin{equation}
    \chi(\omega) = \int_0^{\infty} \dd t e^{-i\omega t} \Tr\left\{\hat{\rho}\comm{\hat{\Lambda}_\mu(t)}{\hat{\Lambda}_\nu}\right\}.
    \label{eq:DefnLinearResponse}
\end{equation}
Analogous formulae can be shown to hold for all QFIM's (that is, for any choice of monotone function $f$). The general relationship is given by,
\begin{equation}
    \mathcal{F}_{\mu\nu}^{\text{unitary}} \{f\} = \frac{2}{\pi}\int_0^{\infty} \dd\omega\,\frac{1-e^{-\beta\omega}}{f(e^{-\beta\omega})}\chi_{\mu\nu}''.
\end{equation}
The connection between quantum geometry and response functions has been exploited in a number of proposed~\cite{bleu2018measuring,klees2020microwave,klees2021ground} and realized~\cite{yu2020experimental,gianfrate2020measurement} experiments.

This connection is particularly significant in light of the relationship between quantum geometry and entanglement, which we discuss further in Sec.~\ref{subsubsec:QFIAndEntanglement}. In particular, the low temperature properties of the entanglement are strongly controlled by the static structure factor, as demonstrated in Ref.~\cite{lambert2019estimates} and Ref.~\cite{menon2023multipartite}.

\subsubsection{The Quantum Variance}\label{sec:QuantumVariance}

In the introduction, we promised that quantum information geometry would offer us a means
by which to separate the classical fluctuations from the quantum fluctuations in a given 
system. The quantum covariance~\cite{frerot2016quantum,malpetti2016quantum} makes this concept explicit by decomposing the covariance between two
operators as,
\begin{equation}
    \text{Cov}_{\hat{\rho}}\{\hat{\Lambda}_{\mu},\hat{\Lambda}_{\nu}\} = 
    \text{Cov}^{\mathcal{Q}}_{\hat{\rho}}\{\hat{\Lambda}_{\mu},\hat{\Lambda}_{\nu}\}
    +
    \text{Cov}^{\mathcal{C}}_{\hat{\rho}}\{\hat{\Lambda}_{\mu},\hat{\Lambda}_{\nu}\}
    \label{eq:CovDecomposition}
\end{equation}
where $\text{Cov}_{\hat{\rho}}^{\mathcal{Q}}\{\cdot,\cdot\}$ is the quantum component of the covariance and $\text{Cov}_{\hat{\rho}}^{\mathcal{C}}\{\cdot,\cdot\}$ is the classical component of the covariance. 

To define the classical covariance, we begin by considering the mixed state defined in Eq.~\ref{eq:DefnQuantumGibbsState}, 
\begin{equation}
    \hat{\rho} = \frac{e^{-\beta\hat{H}}}{\mathcal{Z}}
\end{equation}
where,
\begin{equation}
    \mathcal{Z} = \Tr\{e^{-\beta\hat{H}}\}
\end{equation}
which can be used to describe any mixed state for a suitable choice of $\hat{H}$. Now add to the Hamiltonian the operators $\hat{\vec{\Lambda}}$ which are coupled via the parameters $\hat{\vec{\lambda}}$
\begin{subequations}
\begin{align}
    \hat{\rho}(\vec{\lambda}) &= \frac{e^{-\beta(\hat{H}+\vec{\lambda}\cdot\hat{\vec{\Lambda}})}}{\mathcal{Z}} \\
    \mathcal{Z}(\vec{\lambda}) &= \Tr{e^{-\beta(\hat{H} + \vec{\lambda}\cdot\hat{\vec{\Lambda}})}}.
\end{align}
\end{subequations}
Using the free energy,
\begin{equation}
    F = -\frac{1}{\beta}\ln(\mathcal{Z}(h)).
\end{equation}
we can now define the classical covariance as,
\begin{equation}
    \text{Cov}_{\hat{\rho}}^{\mathcal{C}}\{\hat{\Lambda}_{\mu},\hat{\Lambda}_{\nu}\} 
    = 
    \frac{1}{\beta}\pdv[2]{F}{\lambda_\mu}{\lambda_\nu}
    \label{eq:DefnClassicalCovariance}
\end{equation}
where the second derivative of the free energy can be recognized as the static thermal susceptibility. The quantum covariance is then defined directly by rearranging Eq.~\ref{eq:CovDecomposition},
\begin{equation}
    \text{Cov}_{\hat{\rho}}^{\mathcal{Q}}\{\hat{\Lambda}_{\mu},\hat{\Lambda}_{\nu}\} =
    \text{Cov}_{\hat{\rho}}\{\hat{\Lambda}_{\mu},\hat{\Lambda}_{\nu}\} - \text{Cov}_{\hat{\rho}}^{\mathcal{C}}\{\hat{\Lambda}_{\mu},\hat{\Lambda}_{\nu}\}
    \label{eq:QuantumCovariance}
\end{equation}
To understand the meaning of this definition, let's first consider the case where $\comm{\hat{H}}{\hat{\vec{\Lambda}}} = 0$. From the perspective of information geometry, this kind of perturbation will change the eigenvalues of the density matrix, but not the eigenvectors. In this case, the classical covariance will be precisely the covariance, and the quantum covariance will be zero. 

In the case where the commutator of the Hamiltonian with the perturbation is non-zero. The evaluation of the second derivative in Eq.~\ref{eq:DefnClassicalCovariance} must now be performed by taking a Taylor series expansion of the exponential about the point where $\vec{\lambda}=\vec{0}$. In general the thermal covariance will be strictly less than the total covariance (except at infinite temperature), and the difference is made up for by the quantum fluctuations quantified by the quantum covariance. Notice that the quantum variance is just the diagonal of the quantum covariance matrix,
\begin{equation}
    \text{Var}_{\hat{\rho}}^{\mathcal{Q}}\{\hat{\Lambda}_{\mu}\} 
    \equiv 
    \text{Cov}_{\hat{\rho}}^{\mathcal{Q}}
    \{\hat{\Lambda}_{\mu},\hat{\Lambda}_{\mu}\}
\end{equation}

The physical meaning of the quantum covariance at finite temperature can be gleaned by considering the case of a free particle in equilibrium with a thermal bath. The quantum variance associated with the position operator is then~\cite{frerot2016quantum},
\begin{equation}
    \text{Var}^{\mathcal{Q}}_{\hat{\rho}}(\hat{x};\hat{\rho}) = \frac{1}{24\pi} \lambda_{dB}^2
\end{equation}
where $\lambda_{dB}=\sqrt{\frac{2\pi\hbar^2}{mk_B T}}$ is the \emph{thermal de Broglie wavelength}. It decreases monotonically to zero as the temperature increases, and thus indicates the falling off of quantum fluctuations at finite temperature. 

The quantum covariance appears to be behaving exactly as we would expect the unitary component of the QGT to behave. To better understand the relationship between the two, we recall a special case of the QGT, the Wigner-Yanase-Dyson (WYD) skew information,
~\cite{wigner1997information,wigner1964positive}, 
\begin{equation}
  \mathcal{I}_{\alpha,\hat{\rho}}(\hat{\Lambda}_{\mu},\hat{\Lambda}_{\nu}) =
  -\frac{1}{2}\tr(\comm{\hat{\Lambda}_{\mu}}{\hat{\rho}^\alpha}\comm{\hat{\Lambda}_{\nu}}{\hat{\rho^{1-\alpha}}}).
  \label{eq:DefnWYDInfo}
\end{equation}~
This QGT corresponds to the following choice of monotone 
function~\cite{shitara2016determining}, 
\begin{equation}
  f_\alpha(x) = \alpha(1-\alpha)\frac{(x-1)^2}{(x^\alpha-1)(x^{1-\alpha}-1)}.
  \label{eq:DefnGenFunc}
\end{equation}
It is shown in Ref.~\cite{frerot2016quantum} that the quantum covariance may be written as, 
\begin{equation}
  \text{Cov}^\mathcal{Q}_{\hat{\rho}}(\hat{\Lambda}_{\mu},\hat{\Lambda}_{\nu})
  =\int_0^1\dd\alpha\,\,
  \mathcal{I}_\alpha(\hat{\Lambda};\hat{\rho}).
  \label{eq:WYDInfoQuantumVariance}
\end{equation}
So we see that the quantum variance is the average over a particular family of QGT's.
The diagonal entries of the SLD QFIM are bounded from above and below by the diagonal entries of the quantum covariance (recall that the Berry curvature doesn't have diagonal entries),
\begin{equation}
    4\text{Var}^{\mathcal{Q}}_{\hat{\rho}}(\hat{\Lambda}_\mu) \leq \mathcal{F}_{\mu\mu}\leq 12 \text{Var}^{\mathcal{Q}}_{\hat{\rho}}(\hat{\Lambda}_\mu)
\end{equation}

Using the representation of the WYD information in terms of response functions, Eq.~\ref{eq:WYDInfoQuantumVariance} can be written as~\cite{frerot2019reconstructing},
\begin{equation}
    \text{Var}_{\mathcal{Q}}(\hat{\Lambda};\hat{\rho}) = \hbar \int_{0}^{\infty}\frac{\dd\omega}{\pi }
    L\left(\frac{\beta\hbar\omega}{2}\right) \chi''_{\hat{\Lambda}}(\omega)
    \label{eq:DefnQvInTermsOfResponseFunctions}
\end{equation}
where,
\begin{equation}
     L\left(x\right) = \coth(x) - \frac{1}{x}
\end{equation}
is the Langevin function. We thus see that the quantum variance corresponds directly to the unitary component of the QFIM described in Eq.~\ref{eq:DefnQFIMUnitaryPart}, making explicit the fact that this component corresponds to quantum fluctuations by subtracting off the thermal fluctuations that would give rise to the non-unitary component of the QFIM in Eq.~\ref{eq:NonUnitaryComponent}. 

To understand the difference between the quantum variance and SLD QFIM 
we consider a single qubit in thermal equilibrium with a bath of inverse
temperature $\beta$. We denote explicitly the gap of the qubit with $\Delta$. 
\begin{equation}
  \rho = \frac{1}{\mathcal{Z}}e^{-\beta\Delta\sigma^z}
\end{equation}
It is easy to rewrite the density matrix explicitly as, 
\begin{equation}
  \rho = \frac{1}{2}(\mathbb{I} - \tanh(\qty|\Delta|\beta)\hat{\sigma}^z)
  \label{eq:DefnQubitDM}
\end{equation}
We consider parameterizations of the Hilbert space generated unitarily by the 
magnetization operator in direction $\hat{n}$,
\begin{equation}
  \hat{\Lambda}(\theta,\phi) = \hat{n}\cdot\vec{\sigma} = \cos(\theta)\sin(\phi)\hat{\sigma}^x +
  \sin(\theta)\sin(\phi)\hat{\sigma}^y + \cos(\phi)\hat{\sigma}^z
\end{equation}
Below we calculate the variance, its quantum and thermal components, and the QFI
as a function of $\beta$, $\theta$, and $\phi$. It is helpful for this purpose
to compute the partition function with a source term,
\begin{align}
  \mathcal{Z}(h) &= \tr\{\exp(-\beta(\Delta\hat{\sigma}^z -
  h\hat{\Lambda}(\theta,\phi)))\}  \nonumber 
\end{align}

\begin{figure}[h!]
  \centering
  \includegraphics[width=0.9\textwidth]{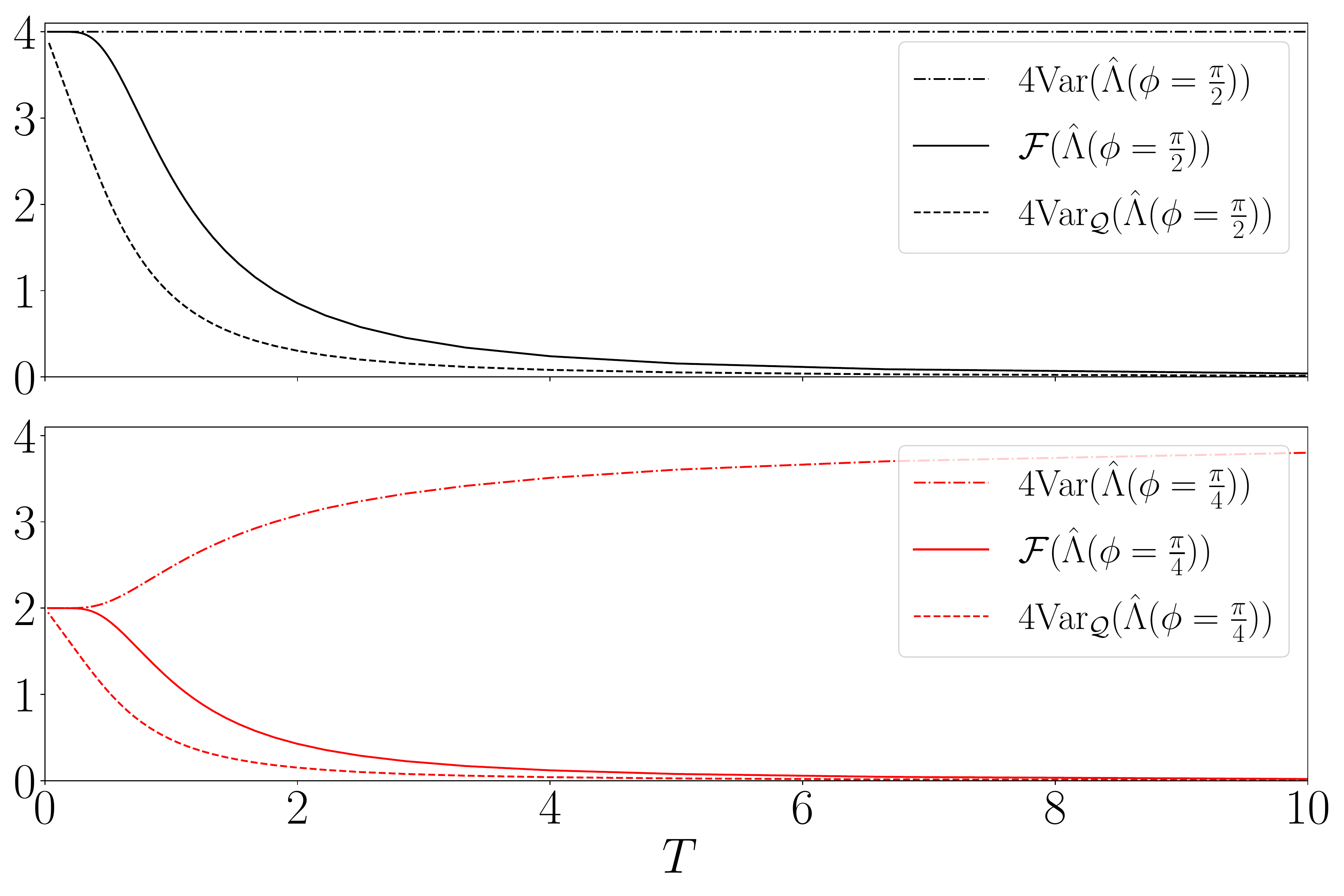}
  \caption[Information geometry of a single qubit]{Plots of the variance, quantum variance, and quantum Fisher
  information for two choices of the magnetization operator.}
  \label{fig:singleQubitFiniteTInfoMagOp}
\end{figure}

The variance, QV (from Eq.~\ref{eq:QuantumCovariance}), and QFI (from Eq.~\ref{eq:DefnQFIMUnitaryPart}), for $\phi=\frac{\pi}{2}$ and $\phi=\frac{\pi}{4}$ are
shown in Fig.~\ref{fig:singleQubitFiniteTInfoMagOp} with details of the calculations given in App.~\ref{App:SampleCalculationSingleQubit}. We can see that regardless of the orientation, the variance will always saturate at its maximal value in the limit of infinite temperature ($\beta\rightarrow 0$).
If we imagine the Bloch sphere with its interior as the prototypical representation of the space of pure and mixed states, then the infinite temperature limit corresponds to the center of this sphere. From an information theory perspective, it is the equivalent of a black hole, as all states taken to this limit become indistinguishable.  This manifests in the QFI going to zero. We see that the quantum variance also tends to zero. While both the QV and QFIM must go to zero in this limit, they do so at slightly different rates. Their difference is not constant but peaks at some value of $\beta$ that scales with the energy gap of the system. 

\subsubsection{Quantum Fisher Information and Entanglement}
\label{subsubsec:QFIAndEntanglement}

Among the most significant aspects of the geometrical structure of quantum state space is its explicit relationship to multipartite which was, as far as we are aware, first made explicit in Ref.~\cite{pezze2009entanglement}, and later elaborated on in Ref.~\cite{hyllus2012fisher,toth2012multipartite}. The diagonal components of the QFIM may be related explicitly to bounds on the amount of multipartite entanglement. These bounds are significant for three reasons. First, they apply even at finite temperature. Second, they are directly accessible through the linear response functions discussion in Sec.~\ref{subsec:relationToLinearResponse}. Third, the detection of entanglement depends in particular on the direction in state space that we choose to examine, thus allowing us to consider a richer structure of entanglement than would be accessible if we only considered the von Neumann entropy. Below we derive the bounds explicitly following the discussion in Ref.~\cite{hyllus2012fisher}. 

Consider an $N$ body system and a Hermitian operator composed of a sum of one-body operators,
\begin{equation}
    \hat{\Lambda} = \sum_{j=1}^N \hat{A}_j
    \label{eq:LocalOperator}
\end{equation}
each with maximum eigenvalue $a_{\text{max}}$, minimum eigenvalue $a_{\text{min}}$. Let $\Delta_a = a_{\text{max}}-a_{\text{min}}$ be the spectral width. The first bound we derive relies only on the component of the QFIM associated with the operator in Eq.~\ref{eq:LocalOperator}, which we hereafter refer to simply as the QFI, $\mathcal{F}$, omitting the indices for simplicity. We also note the following
\begin{enumerate}[label=\alph*]
    \item The set of $k$-producible states is convex. That is to say, the producibility is non-increasing under mixture.
    \item The Fisher information is convex under mixture,
    $\mathcal{F}\{p\hat{\rho}_1 + (1-p)\hat{\rho}_2\}\leq p\mathcal{F}\{\hat{\rho}_1\} + (1-p)\mathcal{F}\{\hat{\rho}_2\} $ for $p\in[0,1]$
    \item For a $N$ body state $\ket{\Psi}=\ket{\psi_A}\otimes\ket{\psi_B}$
    we have $\text{Var}_{\ket{\Psi}}(\hat{\Lambda}) = \text{Var}_{\ket{\psi_A}}(\hat{\Lambda}^{(A)}) + \text{Var}_{\ket{\psi_B}}(\hat{\Lambda}^{(B)})$
    \item For a $N$ body pure state $\ket{\Psi}$, $4\text{Var}_{\ket{\Psi}}(\hat{\Lambda}) \leq \Delta_a N^2$ with the maximum achieved by the GHZ state,
    \begin{equation}
        \ket{\text{GHZ}_N} = \frac{1}{\sqrt{2}}\left(\ket{a_{\text{min}}}^{\otimes N}+\ket{a_{\text{max}}}^{\otimes N}\right) 
    \end{equation}
\end{enumerate}
By (a) and (b) the maximal QFI will occur for pure states, so we limit our discussion to that case.
Now we consider the $k$ producible state,
\begin{equation}
    \ket{\Psi} = \bigotimes_{\ell = 1}^M \ket{\psi_{\ell}}
\end{equation}
where each factor $\ket{\psi_{\ell}}$ contains $N_{\ell}$ particles and the largest $N_l$ is equal to $k$. For a given value of $k$, the largest possible QFI is achieved by a state with $s=\lfloor\frac{N}{k} \rfloor$ factors in a $\ket{\text{GHZ}_k}$ state with the remaining particles in a $\ket{\text{GHZ}_{r}}$ state where $r = N-sk$. The QFI for a $k$-producible state is thus bounded from above,
\begin{equation}
    \mathcal{F} \leq \Delta_a(sk^2 + r^2)
    \label{eq:ElementBoundOnQFI}
\end{equation}
with the bound for separable states given by~\cite{pezze2009entanglement},
\begin{equation}
    \mathcal{F}(\hat{\rho}_{\text{separable}}) \leq N 
    \label{eq:BoundOnQFIElementSeparableStates}
\end{equation}

In Refs.~\cite{hyllus2012fisher,toth2012multipartite}, a second inequality is derived, which is aimed specifically at systems whose degrees of freedom can be represented in terms of the spin. Let's consider the case where the local operators in Eq.~\ref{eq:LocalOperator} are given by,
\begin{equation}
    A_j = \mathbf{n}_j\cdot\mathbf{S}_j
\end{equation}
where $\mathbf{n}_j=(\cos(\theta_j)\sin(\phi_j),\sin(\theta_j)\sin(\phi_j),\cos(\phi_j))$ where $\theta\in[0,2\pi)$ and $\phi\in[0,\pi)$. Let us consider the case where the orientation on each site is the same by dropping the subscript $j$ on the unit vector. We denote the corresponding generator, $\hat{\Lambda}_{\mathbf{n}}$ and the corresponding QFI $\mathcal{F}_{\mathbf{n}}$. We can now consider the direction averaged QFI,
\begin{equation}
    \bar{\mathcal{F}} = \int P(\mathbf{n}) \mathcal{F}_{\mathbf{n}} \sin(\phi)\dd\theta\dd\phi
\end{equation}
where $P(\mathbf{n})$ is a function such that $1 = \int P(\mathbf{n}) \sin(\phi)\dd\theta\dd\phi$. If we choose the case of $P(\mathbf{n}) = \delta(\mathbf{n} - \mathbf{a})$ we find the QFI for the particular orientation $\mathbf{a}$ of the generator and the inequality in Eq.~\ref{eq:ElementBoundOnQFI} would apply. Instead, we consider the uniform case where $P(\mathbf{n}) = \frac{1}{4\pi}$, for which
\begin{equation}
    \bar{\mathcal{F}} = \frac{\mathcal{F}_{\mathbf{x}} + \mathcal{F}_{\mathbf{y}} + \mathcal{F}_{\mathbf{z}}}{3}
\end{equation}
In Ref.~\cite{hyllus2012fisher} the above expression is given the following bound for $k$-producible states,
\begin{equation}
    \bar{\mathcal{F}}(\hat{\rho}_{\text{k-producible}}) = \frac{1}{3}\left(s \left(k^2 + 2k - \delta_{k,1}\right) + r^2 + 2r -\delta_{r,1}\right)
    \label{eq:MeanBoundQFI}
\end{equation}
where $s=\lfloor\frac{N}{k} \rfloor$ and $r = N-sk$ are defined as earlier. For fully separable states the above bound reduces to,
\begin{equation}
    \bar{\mathcal{F}}(\hat{\rho}_{\text{separable}}) = \frac{2}{3}N
    \label{eq:BoundOnMeanQFISeparableStates}
\end{equation}
To illustrate the applications of Eq.~\ref{eq:ElementBoundOnQFI} and Eq.~\ref{eq:BoundOnMeanQFISeparableStates} we consider the case of $2$ partite states. In the following discussion, we consider the QFI density, which we indicate with $f=\frac{\mathcal{F}}{N}$ or $\bar{f}=\frac{\bar{\mathcal{F}}}{N}$.

\begin{figure}[t!]
    \centering
    \includegraphics[width=0.65\textwidth]{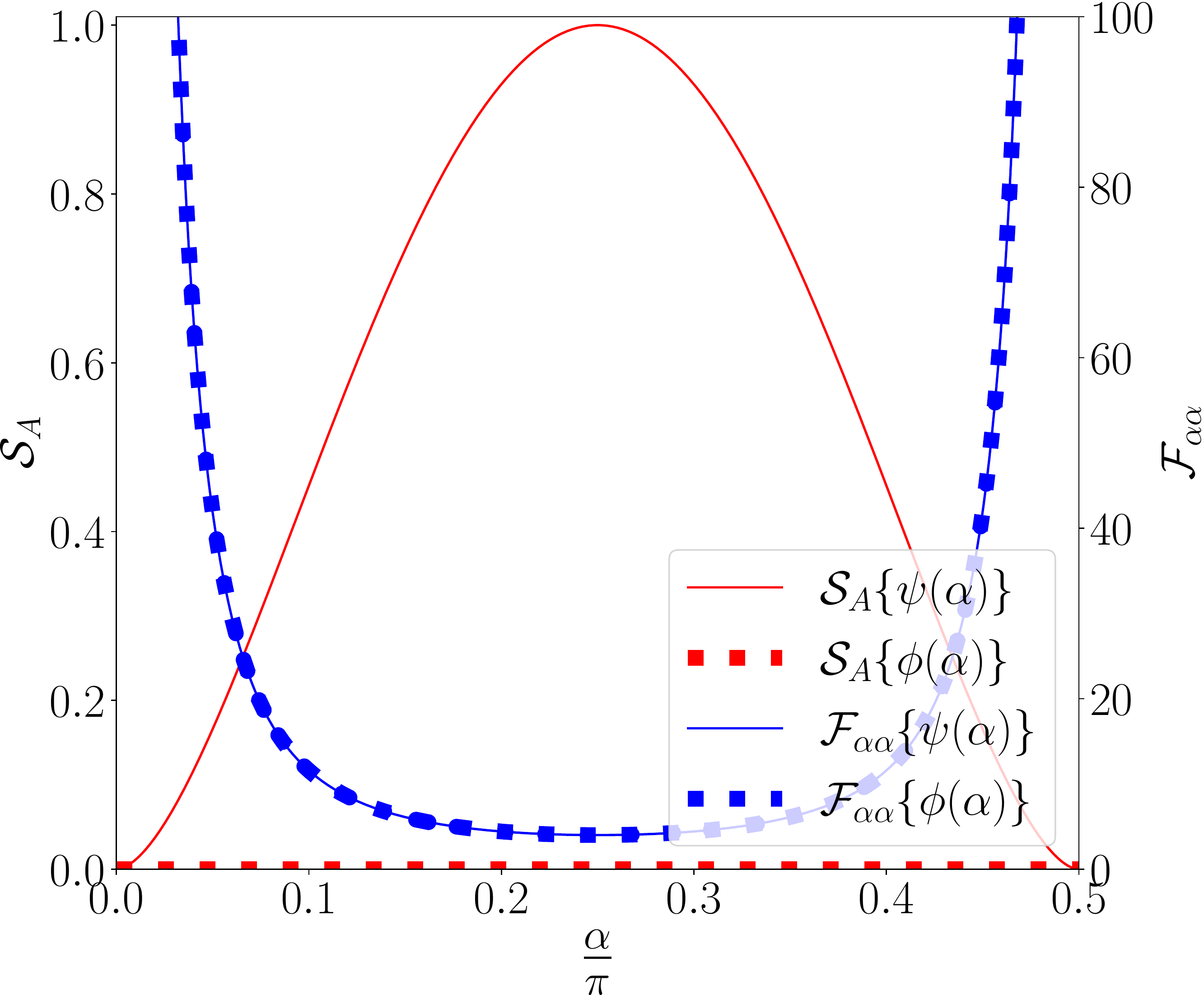}
    \caption[Geometry of interpolated entangled and separable states]{Fisher information and Entanglement entropy of the states   in Eq.~\ref{eq:BellState} and Eq.~\ref{eq:ProductState}. Here the Fisher information is taken with respect to the parameter $\lambda$ itself, and is identical for both states. The EE for $\ket{\psi}$ attains a maximum of one bit when the state is maximally entangled, while the EE for $\ket{\phi}$ is always zero.}
    \label{fig:BellStateEntanglementEntropy}
\end{figure}

The extent to which the above inequalities detect entanglement depend sensitively on the choice of the operator $\hat{\Lambda}$ in Eq.~\ref{eq:LocalOperator}, or equivalently, on the direction in state space in which we choose to perturb the state. To see this,
consider the following two states,
\begin{subequations}
\begin{align}
    \ket{\psi(\alpha)} &= \cos(\alpha)\ket{\uparrow_A\downarrow_B} - \sin(\alpha)\ket{\downarrow_A\uparrow_B},
    \label{eq:BellState}  \\
    \ket{\phi(\alpha)} &= \cos(\alpha)\ket{\uparrow_A\downarrow_B} +
    \sin(\alpha)\ket{\downarrow_A\downarrow_B}
    \label{eq:ProductState}
\end{align}
\end{subequations}
on the interval $\alpha\in [0,\frac{\pi}{2}]$. At $\alpha=\frac{\pi}{4}$ the first state is the singlet, which we know is maximally entangled. This state is maximally entangled, containing one bit of entanglement entropy, as can be seen from the solid red curve in Fig.~\ref{fig:BellStateEntanglementEntropy}.
The second state is separable for all possible values of $\lambda$ and therefor has an entanglement entropy of zero. For both states, we can compute the QFI associated with the parameter $\lambda$ and find that they are identical. The QFI associated with this parameter tells us nothing about the entanglement content of the state. It's easy to see that any single parameter family of states that interpolates smoothly between two orthogonal states in the form given about will always yield the QFI,
\begin{equation}
    \mathcal{F}_{\alpha\alpha} = \sec^2(\alpha)\csc^2(\alpha)
    \label{eq:QfiPureStateInterpolation}
\end{equation}

To distinguish between entangled and separable spin states, we must consider a path generated by an operator that is local in the spin degrees of freedom. Even for two spins, there is considerable freedom in which operators we might choose. Two natural choices are the ferromagnetic and anti-ferromagnetic combinations of each of the three spin operators,
\begin{equation}
    \hat{\Lambda}^{\pm}_{\mu} = \hat{S}_A^{\mu} \pm \hat{S}_B^{\mu}
    \label{eq:TwoSpinGenerators}
\end{equation}
where $\mu\in\{x,y,z\}$.

\begin{figure}[h!]
    \centering
    \includegraphics[width=0.8\textwidth]{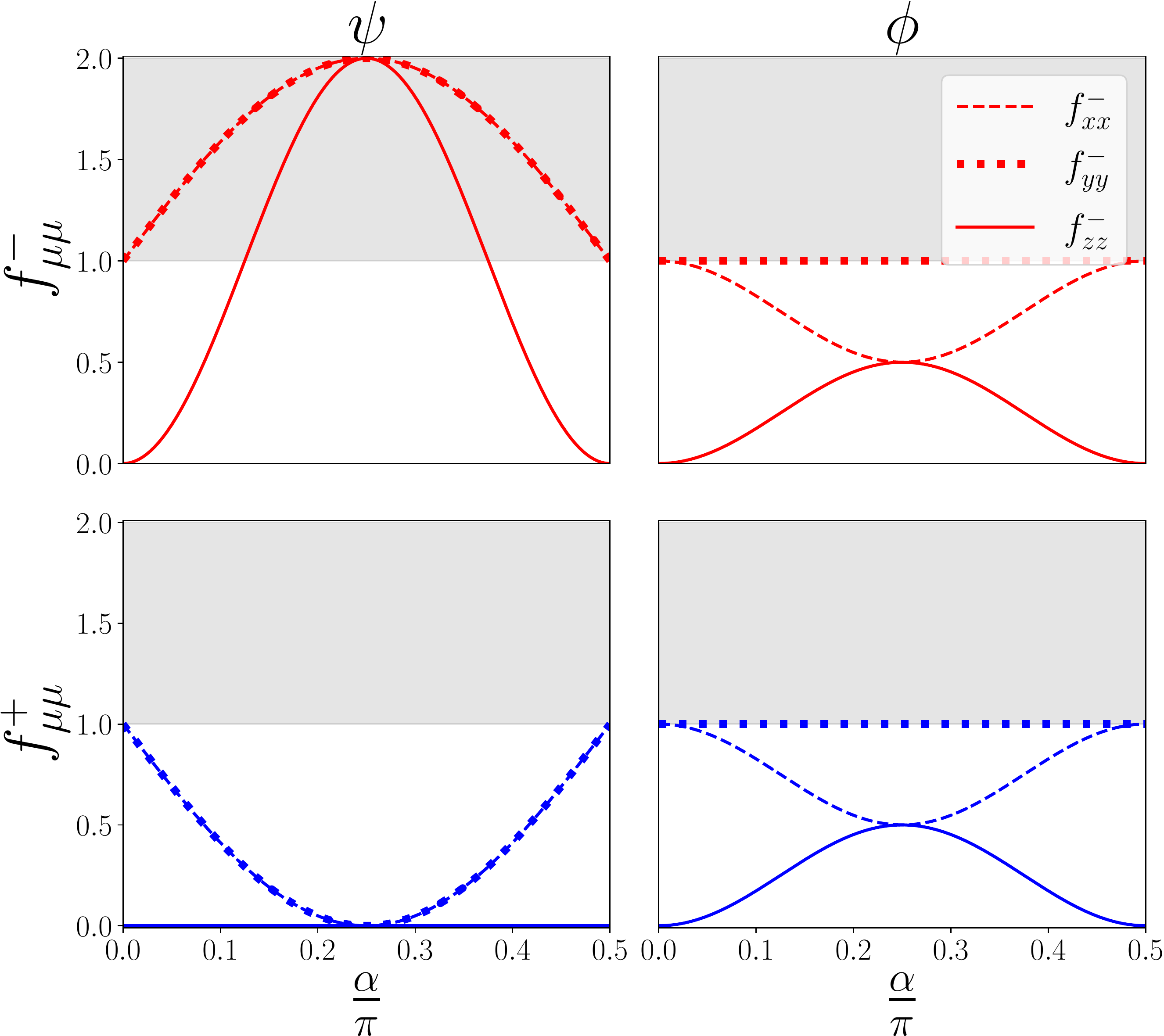}
    \caption[Quantum Fisher information for entangled and separable states]{Diagonal entries of the QFIM matrix associated with the three ferromagnetic (blue) and anti-ferromagnetic (red) generators defined in Eq.~\ref{eq:TwoSpinGenerators}. The three line styles correspond to the $x$ (dashed), $y$ (dotted), and $z$ (solid) lab frame directions. The gray region indicates the bound given in Eq.~\ref{eq:BoundOnMeanQFISeparableStates}.}
    \label{fig:qfiDensitiesTwoSpins}
\end{figure}

Each of these generators defines a unitary operator, which induces an infinitesimal displacement in the state along a path parameterized by $\lambda$,
\begin{equation}
    \ket{\alpha,\dd \lambda_{\mu}} = e^{i\dd \lambda_{\mu} \Lambda^{\pm}_{\mu}}\ket{\alpha}
\end{equation}
The distance between $\ket{\lambda,0}$ and $\ket{\lambda,\dd h}$ is then given by (following Eq.~\ref{eq:DefnPureStateQFIM}),
\begin{equation}
    \mathcal{F}_{\mu\mu}^{\pm}\{\psi\} =4\text{Cov}_{\psi}(\hat{\Lambda}_\mu^{\pm},\hat{\Lambda}_{\mu}^{\pm})
\end{equation}
The QFI densities generated by the ferromagnetic and anti-ferromagnetic operators are give in Fig.~\ref{fig:qfiDensitiesTwoSpins}. 

While the separable state, $\ket{\phi}$ never exceeds the bound given in Eq.~\ref{eq:BoundOnQFIElementSeparableStates}, we can see the for the entangled state $\ket{\psi}$
the bound is exceeded by three of the different paths chosen. First, for the path given in terms of the $z$ component for the anti-ferromagnetic spin combination, bipartite entanglement is detected for a window the parameter space $\lambda\in\left[\frac{\pi}{8},\frac{3\pi}{8}\right]$. The bound is also always violated for the parameterizations generated by the anti-ferromagnetic $x$ and $y$ operators for the entire span of $\lambda$ except at the separable points. 

\begin{figure}
    \centering
    \includegraphics[width=1.\textwidth]{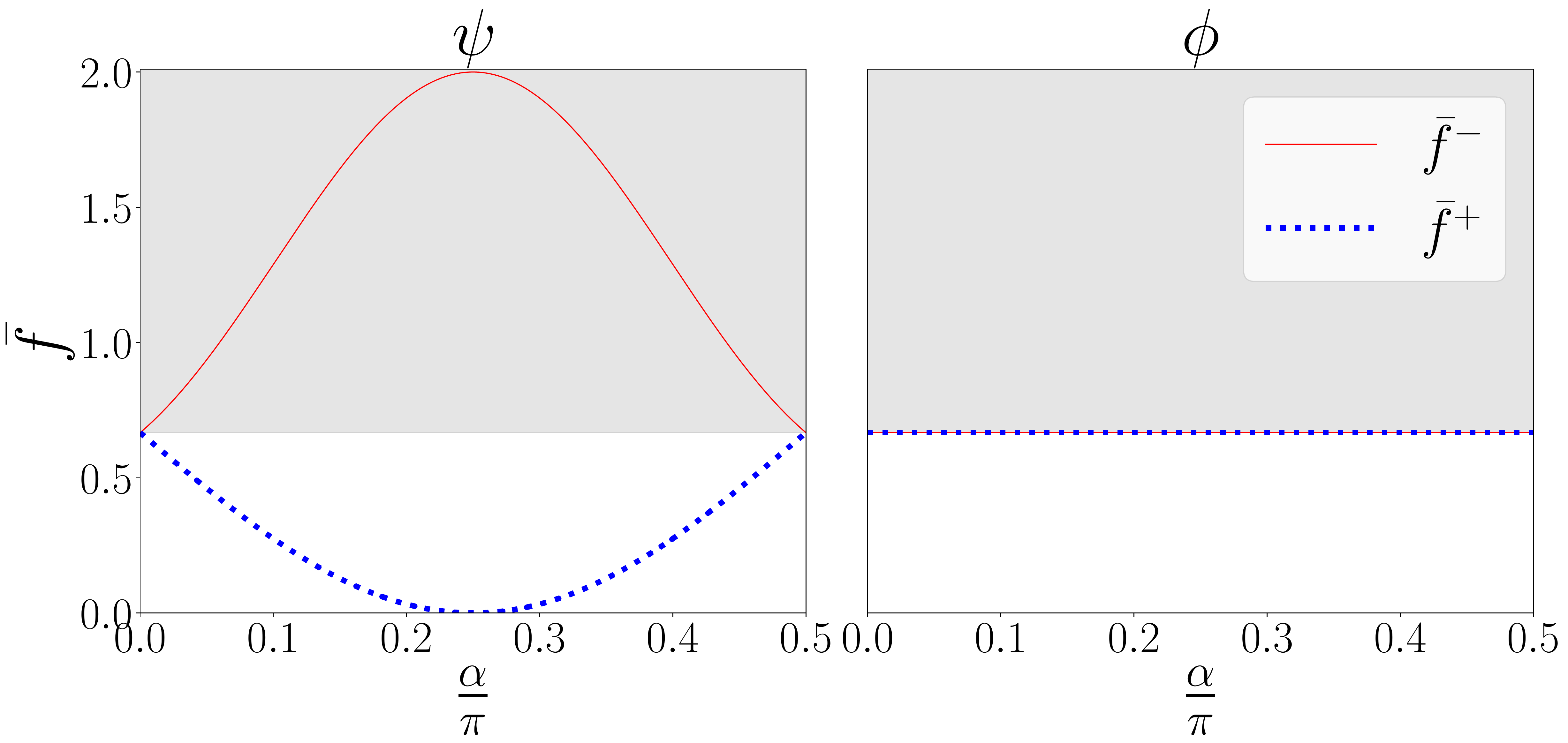}
    \caption[QFIM trace for separable and pure states]{Trace of the QFIM matrix for the three symmetric (blue) and antisymmetric (red) generators defined in Eq.~\ref{eq:TwoSpinGenerators}. The gray region indicates the bound on entanglement given in Eq.~\ref{eq:BoundOnMeanQFISeparableStates} for detecting $2$-partite entanglement. Notice that only the symmetric choice detects the bipartite entanglement in the state $\ket{\psi}$.}
    \label{fig:meanQfiDenistiesTwoSpins}
\end{figure}

The three ferromagnetic operators defined in Eq.~\ref{eq:TwoSpinGenerators} (or, equivalently, the three anti-ferromagnetic spin operators), are related to each other by global rotations. We can thus consider the QFIM corresponding to the three-dimensional submanifold whose parameterization is generated by these operators and take its trace. The results for the ferromagnetic and anti-ferromagnetic operators are given in Fig.~\ref{fig:meanQfiDenistiesTwoSpins}. The bound corresponding to Eq.~\ref{eq:BoundOnMeanQFISeparableStates} is again given in gray. Here we see that only the anti-ferromagnetic spin combinations detect entanglement. 

Notice that at the point, $\alpha = \frac{\pi}{4}$ all three components of the QFIM for the ferromagnetic combination are zero. The singlet is an eigenstate of the magnetization operator, having total magnetization zero in any of the three spin directions. This special case implies a general principle. If a state $\ket{\psi}$ is an eigenstate of the generator of a parameterization, then its QFI along that parameterization will be zero. This is easy to see if one considers the fact that the QFI for a pure state is just the covariance of the generator for that state. 

More generally, if the Hamiltonian of a system commutes with the generator, then the ground state of that Hamiltonian will tend to exhibit zero QFI. An exception is in the case where the ground state is degenerate. For example, the ferromagnetic Heisenberg chain has a three-fold degenerate ground state manifold, which can be used to construct entangled states for which the QFI associated with the ferromagnetic generator does detect entanglement. 

\subsubsection{Fidelity Susceptibility}\label{subsubsec:FidelitySusceptibility}

The geometrical structure of the Hilbert space introduced in Sec.~\ref{sec:TheoryOfInformationGeometry} applies to any parameterization of the state space. One case of particular physical interest is when that parameterization is generated by considering the ground state of the Hamiltonian,
\begin{equation}
    \hat{H}(\vec{\Lambda}) = H_0 + \vec{\lambda}\cdot\hat{\vec{\Lambda}}
    \label{eq:ParameterizedHamiltonian}
\end{equation}
where the parameters $\vec{\lambda}$ can be used to drive phase transitions in the ground state manifold. In this case, the QFIM is called the \emph{fidelity susceptibility}~\cite{zanardi2006ground,gu2008fidelity,gu2010fidelity} (up to a conventional factor of four, which we omit). It is also possible to consider the fidelity susceptibility with respect to parameters that are not driving the phase transition, such as a twist in the boundary conditions~\cite{Laflorencie2001,Thesberg2011,Thesberg2014}, in which case it would be more appropriate to associate it with the QFIM.

To construct the QGT explicitly, let $\ket{\Psi_0(\lambda)}$ be the ground state of Eq.~\ref{eq:ParameterizedHamiltonian} (we consider one component for simplicity). The infinitesimally neighboring state $\ket{\Psi(\lambda+\dd\lambda)}$ can be found by applying perturbation theory,
\begin{equation}
    \ket{\Psi_0(\lambda_{\mu}+\dd\lambda_{\mu})} =\ket{\Psi_0(\lambda_{\mu})} 
    + \dd\lambda_{\mu}\sum_{n\neq 0} \frac{\bra{\Psi_n}\hat{\Lambda}_{\mu}\ket{\Psi_0}\ket{\Psi_n}}{E_n-E_0}
\end{equation}
where, for simplicity, we consider one parameter at a time. 
Then, applying Eq.~\ref{eq:QuantumGeometricTensor} and taking the real part, we recognize the QGT as,
\begin{equation}
    \tilde{\mathcal{Q}}_{\mu\nu} = \sum_{n\neq 0} \frac{\bra{\Psi_0}\hat{\Lambda}_{\mu}\ket{\Psi_n}\bra{\Psi_n}\hat{\Lambda}_{\nu}\ket{\Psi_0}}{(E_n-E_0)^2}.
    \label{eq:FidelityQGTPeturbation}
\end{equation}
Provided that we avoid level crossings, the change in the parameters $\lambda$ can be represented by the unitary operator~\cite{zanardi2007information},
\begin{equation}
    \hat{U} = \sum_{n}\ket{\Psi(\vec{\lambda}+\dd\vec{\lambda})}
    \bra{\Psi(\vec{\lambda})}
\end{equation}
We may identify the associated Hermitian operator via,
\begin{equation}
    \hat{\tilde{\Lambda}}_{\mu} = i(\partial_{\mu}\hat{U})\hat{U}
    \label{eq:FSGenerator}
\end{equation}
and rewrite the QGT in the covariance form,
\begin{equation}
    \tilde{\mathcal{Q}}_{\mu\nu} = \text{Cov}_{\Psi_0}\left\{\hat{\tilde{\Lambda}}_{\mu},\hat{\tilde{\Lambda}}_{\nu}\right\}
    \label{eq:FidelityQGT}
\end{equation}
We denote the above QGT with a tilde in order to make a distinction between two different but related approaches to quantum metrology. Both approaches involve first defining a Hermitian operator $\hat{\Lambda}$. In the first approach, the associated unitary $\exp{-i\lambda\hat{\Lambda}}$ is applied directly to our initial state. The resulting QGT is denoted $\mathcal{Q}_{\mu\nu}$, where
\begin{align}
\mathcal{Q}_{\mu\nu} 
   &= 
    \bra{\Psi_0}\hat{\Lambda}_{\mu}\hat{\Lambda}_{\nu}\ket{\Psi_0} - \bra{\Psi_0}\hat{\Lambda}_{\mu}\ket{\Psi_0}\bra{\Psi_0}\hat{\Lambda}_{\nu}\ket{\Psi_0}\nonumber\\
   &= \text{Cov}_{\Psi_0}\left\{\hat{\Lambda}_{\mu},\hat{\Lambda}_{\nu}\right\}.
\end{align}
In the second approach, the initial state is taken to be the ground state of a Hamiltonian to which $\hat{\Lambda}$ is applied as a perturbation. This is the QGT denoted $\tilde{\mathcal{Q}}_{\mu\nu}$ in Eq.~\ref{eq:FidelityQGT}. The two are in fact related by a series of inequalities~\cite{zanardi2007information} which we reproduce here for completeness.

Let $\epsilon_{nm} = E_n-E_m$ and let $\epsilon_{10}$ be the smallest spectral gap. Now we may rewrite Eq.~\ref{eq:FidelityQGTPeturbation} as
\begin{align}
    \tilde{\mathcal{Q}}_{\mu\nu} &= \sum_{n\neq 0} \frac{\bra{\Psi_0}\hat{\Lambda}_{\mu}\ket{\Psi_n}\bra{\Psi_n}\hat{\Lambda}_{\nu}\ket{\Psi_0}}{\epsilon_{n0}^2} \nonumber \\
    &= \frac{1}{\epsilon_{n'0}^2} \sum_{n\neq 0} 
    \frac{\epsilon_{10}^2}{\epsilon_{n0}^2}\bra{\Psi_0}\hat{\Lambda}_{\mu}\ket{\Psi_n}\bra{\Psi_n}\hat{\Lambda}_{\nu}\ket{\Psi_0}
    \nonumber \\
    &\leq 
    \frac{1}{\epsilon_{10}^2}\sum_{n\neq 0} \bra{\Psi_0}\hat{\Lambda}_{\mu}\ket{\Psi_n}\bra{\Psi_n}\hat{\Lambda}_{\nu}\ket{\Psi_0}
    \nonumber \\
    &=\frac{1}{\epsilon^2_{10}}
    \left(\sum_n \bra{\Psi_0}\hat{\Lambda}_{\mu}\ket{\Psi_n}\bra{\Psi_n}\hat{\Lambda}_{\nu}\ket{\Psi_0} - \bra{\Psi_0}\hat{\Lambda}_{\mu}\ket{\Psi_0}\bra{\Psi_0}\hat{\Lambda}_{\nu}\ket{\Psi}_0\right)
    \nonumber \\
    &= \frac{1}{\epsilon^2_{10}}\mathcal{Q}_{\mu\nu}
\end{align}
Thus we find that, $\tilde{\mathcal{Q}}_{\mu\nu} \leq \frac{1}{\epsilon_{10}^2}\mathcal{Q}_{\mu\nu}$.

At a level crossing critical point, we expect the QGT to diverge. The nature of this divergence was determined in Ref.~\cite{venuti2007quantum,Schwandt2009,Albuquerque2010}. In order to analyze the scaling of the QGT it is useful to consider the \emph{QGT density},
\begin{equation}
    \tilde{q}_{\mu\nu} = L^{-d}\tilde{\mathcal{Q}}_{\mu\nu}
    \label{eq:IntensiveQGT}
\end{equation}
where $d$ is the spatial dimension of the system.
Following Ref.~\cite{venuti2007quantum}, we consider a scale transform $x\rightarrow \alpha x$ and $\tau \rightarrow \alpha^{\zeta}\tau$. Here $\zeta$ is the dynamical critical exponent. Let $\Delta_{\mu}$ be the scaling dimension of the operator $\hat{\Lambda}_{\mu}$. If $\lambda_c$ is the critical point, the correlation length will diverge as $\xi \sim \qty|\lambda-\lambda_c|^{-\nu}$.
In the region approaching the critical point the QGT density will diverge as,
\begin{equation}
    \tilde{q}_{\mu\nu} \sim \qty|\lambda-\lambda_c|^{\nu\left(\Delta_{\mu} + \Delta_{\nu} -2\zeta - d\right)}
    \label{eq:FSFieldScaling}
\end{equation}
This also leads to a finite size scaling relation at the critical point itself where,
\begin{equation}
    \tilde{q}_{\mu\nu} \sim L^{-(\Delta_{\mu}+\Delta_{\nu}-2\zeta-d)}.
    \label{eq:FSSizeScaling}
\end{equation}
If the operator $\hat{\Lambda}_{\mu}$ drives the transition, as is the case when considering the fidelity susceptibility, then $\Delta_\mu=d+\zeta-1/\nu$ and the diagonal terms in Eq. \ref{eq:FSSizeScaling} become~\cite{Schwandt2009,Albuquerque2010}:
\begin{equation}
    \tilde{q}\sim L^{2/\nu-d}.
    \label{eq:FSSizeScaling2}
\end{equation}

These scaling relations can be contrasted with the scaling of the QGT for a unitary perturbation generated directly by the operator $\hat{\Lambda}$. In this case the dynamical critical exponent doesn't enter in, and we instead have,
\begin{equation}
    q_{\mu\nu} \sim \qty|\lambda-\lambda_c|^{\nu(\Delta_{\mu}+\Delta_{\nu} -d)}
    \label{eq:FIFieldScaling}
\end{equation}
and at the critical point itself the finite size scaling,
\begin{equation}
    q_{\mu\nu} \sim  L^{-(\Delta_{\mu}+\Delta_{\nu}-d)}.
    \label{eq:FISizeScaling}
\end{equation}
As above, if we assume $\Delta_\mu=\Delta_\nu=2+\zeta-1/\nu$, we obtain for the diagonal term:
\begin{equation}
    q \sim  L^{2/\nu-d-2\zeta}.
    \label{eq:FISizeScaling2}
\end{equation}
For comparison, it is useful to note that the closely related energy susceptibility, $\chi^e=-\partial^2e_0/\partial \lambda^2$, with $e_0$ the ground-state energy per site, scales as
\begin{equation}
    \chi^e \sim  L^{2/\nu-d-\zeta}.
    \label{eq:e0Scaling}
\end{equation}
This result follows immediately from the scaling of the critical free energy density, $f_s\sim\qty |\lambda-\lambda_c|^{\nu(d+\zeta)}$, with $\chi^e$ the second derivative of $f_s$.
Notice that the dynamical critical exponent appears in Eq.~\ref{eq:FSFieldScaling} and Eq.~\ref{eq:FSSizeScaling} but not in Eq.~\ref{eq:FIFieldScaling} and Eq.~\ref{eq:FISizeScaling}. This is due to the fact that the two metrics consider different parameterizations. In particular, the metric $\tilde{\mathcal{Q}}_{\mu\nu}$ determines the distance between the ground state of a Hamiltonian at some value of the Hamiltonian parameter $\lambda_{\mu}$ and some nearby value $\lambda_{\mu}+\dd\lambda_{\mu}$, where $\lambda_{\mu}$ couples to the operator $\hat{\Lambda}_{\mu}$. The metric $\mathcal{Q}_{\mu\nu}$, however, determines the distance between a state $\ket{\psi}$ and the infinitesimally neighboring state $\exp{-i\dd\lambda_{\mu}\hat{\Lambda}_{\mu}}\ket{\psi}$. As we describe in Eq.~\ref{eq:FSGenerator}, the perturbation to the Hamiltonian may be recast as a unitary transformation on the ground state provided that the change in parameters occurs adiabatically. The contribution of the dynamical critical exponent in the Fidelity susceptibility reflects the necessity of the transformation being adiabatic if we are to keep ourselves in the ground state. This difference in scaling turns out to have important implications for quantum sensing~\cite{rams2018limits}. 

The above scaling discussion coupled with the relationship between the QFIM and genuine multipartite entanglement implies that as long as $\hat{\Lambda}$ is relevant in the sense of the renormalization group, the associated entanglement will diverge. In particular, if one had knowledge of both the dynamical critical exponent and the fidelity susceptibility, one could determine the scaling of a lower bound on the quantum Fisher information associated with the operator $\hat{\Lambda}$ and hence the multipartite entanglement. Reciprocally, if one knew both the scaling of the quantum Fisher information and the dynamical critical exponent, one could compute the scaling of an upper bound on the associated fidelity susceptibility. 

\subsubsection{Quantum Sensing}~\label{subsubsec:quantumSensing}

Quantum sensing is the utilization of quantum resources, such as entanglement, to enhance the precision and sensitivity of measurements~\cite{giovannetti2006quantum,giovannetti2011advances}. The example of the weather stations describing in Sec.~\ref{sec:PhyiscsOfInformationGeometry} can be mapped directly onto a general quantum framework, whereby a probe is made to interact with a system for some amount of time, during which the probe acquires a parameter dependent phase~\cite{pezze2014quantum}. The probe is then measured and the phase estimated. 
The maximum precision with which the parameter can be estimated is given by the Cram\'{e}r-Rao bound,
\begin{equation}
    \text{Cov}_P\{\tilde{\lambda}_{\mu},\tilde{\lambda}_{\nu}\}
    \geq  
    \frac{1}{4F_{\mu\nu}}.
\end{equation}
which we derived in Sec.~\ref{subsubsec:FisherInformation}.
For an extensive review of quantum sensing, see Ref.~\cite{degen2017quantum} as well as Ref.~\cite{toth2014quantum}.

In Ref.~\cite{toth2020activating}, the authors introduce a bound on the metrological usefulness of a state $\hat{\rho}$ for a parameterization generated by the operator $\hat{\Lambda}$. In particular, if $g_{\lambda\lambda}^{\text{separable}}$ is a component of the metric generated by $\hat{\Lambda}$ optimized over all separable states, then the metrological gain, $\mathcal{G}$ can be quantified as,
\begin{equation}
    \mathcal{G}\{\hat{\Lambda}\} \coloneqq \frac{g_{\lambda\lambda}(\hat{\rho})}{g_{\lambda\lambda}^{\text{separable}}}.
\end{equation}
For any state with $\mathcal{G}\geq 1$ there is some metrologically useful entanglement. 

The bounds in Eq.~\ref{eq:BoundOnQFIElementSeparableStates} and Eq.~\ref{eq:BoundOnMeanQFISeparableStates} imply that the sensitivity of a probe to quantum measurements is enhanced by the presence of entanglement. This fact can be used to exploit quantum critical regions for increased measurement sensitivity, as discussed in Ref.~\cite{frerot2018quantum}. As mentioned in Sec.~\ref{subsubsec:FidelitySusceptibility}, the difference in the scaling relations for parameterizations controlled adiabatically by a parent Hamiltonian, and those generated directly by a bounded unitary operator lead to different scaling in measurement sensitivity at the critical point~\cite{rams2018limits}. Topological spins chains~\cite{pezze2017multipartite} and atomic ensembles have also been proposed as useful platforms for quantum sensing due to their geometrical properties~\cite{pezze2018quantum}.

Recent work has also examined the applications of the QFIM to the detection of periodically oscillating magnetic fields, using spin chains as intermediates~\cite{mishra2021driving}. Recently, the dynamical phase corresponding to a time crystal has also been shown to have a robust quantum advantage in measurement sensitivity.

\subsubsection{Chern Number and the Euler Characteristic}
\label{subsubsec:ChernNumberAndTopo}

Topology is playing an increasingly central role in the study of condensed matter and many body systems~\cite{hasan2010colloquium,sato2017topological,rachel2018interacting}. Quantum information geometry facilitates a link between the geometry of state space manifolds, response function, and the topological invariants that characterize those systems. A recent achievement was the classification of $N$ level fermionic and bosonic systems according to a generalization of the Bloch-sphere~\cite{kemp2021nested}. Another 

Studies have found that the Euler Characteristic in the first Brillouin zone is also indicative of topologically non-trivial phases~\cite{ma2013euler,zhang2022revealing}. In this line of research, one considers the topology and geometry of the single particle states in a system that is well described by band theory. Recently such systems have been studied in the context of quantum sensing, see for example~\cite{sarkar2022free}. An especially important result in this context is given by Ref.~\cite{ozawa2021relations} which bounds the volume of a two-dimensional parameter manifold (see Eq.~\ref{eq:Volume}) with the Chern number (see Eq.~\ref{eq:ChernNumber}),
\begin{equation}
    \text{Vol}_g \geq \pi |C|.  
\end{equation}

The quantum geometric tensor associated with non-trivial band structures has also been measured experimentally using the relationship between the quantum geometric tensor and the response function discussed in Sec.~\ref{subsec:relationToLinearResponse} (see Ref.~\cite{gianfrate2020measurement}).

\subsubsection{State Space Curvature}
\label{subsubsec:StateSpaceCurvature}

The curvature associated with different many-body phases has also been examined in several contexts~\cite{zanardi2007bures,lambert2022state,erdmenger2020information}. While the physical implications of the state space curvature are not yet fully understood, it has been shown to clearly indicate quantum criticality, including in systems that lack a local order parameter~\cite{lambert2020revealing}. In Ref.~\cite{zanardi2007bures} it was shown that the state space curvature--with inverse temperature as one parameter, and the driving parameter of a quantum phase transition as the other--detects different scaling regimes in the vicinity of the critical point.

In Ref.~\cite{lambert2022state} some progress was made in interpreting the state space curvature through its apparent relationship to multipartite entanglement. By considering spherical slices of state space generated by perturbing the Hamiltonian with the staggered magnetization operator,
\begin{equation}
    \hat{\Lambda}_{\mathbf{n}(\theta,\phi)} = \sum_r (-1)^r \mathbf{n}(\theta,\phi)\cdot\mathbf{S}.
\end{equation}
where $\hat{n}=(\cos(\theta)\sin(\phi),\sin(\theta)\sin(\phi),\cos(\phi))$, coupled to a source Hamiltonian via the field $h$,
\begin{equation}
    H = H_0 + h\hat{\Lambda}_{\mathbf{n}(\theta,\phi)}
    \label{eq:SourceHamiltonian}
\end{equation}
The size of these manifolds can then be computed using Eq.~\ref{eq:Volume},
\begin{equation}
    Vol_{\psi} = \int_{\Omega} \sqrt{\det(\mathcal{F}_{\mu\nu} \left\{\psi(\theta,\phi)\right\})}\dd \theta \dd\phi
\end{equation}
where $\mathcal{F}_{\mu\nu}$ is the metric for the sphere with tangent operators,
\begin{align}
    \hat{\Lambda}_{\theta} &= \sum_r (-1)^r \partial_{\theta} \mathbf{n}\cdot\mathbf{S}_r \\
    \hat{\Lambda}_{\phi} &= \sum_r (-1)^r \partial_{\phi} \mathbf{n}\cdot\mathbf{S}_r
\end{align}
and the curvature is estimated by comparing the scaling of this volume as a function of the coupling field to the scaling of the volume (surface area) of a sphere in flat space. For the $2$-manifold parameterized by the orientation of $\mathbf{n}$, we expect the volume to scale as,
\begin{equation}
    V_{\text{flat}} = 4\pi (ah)^2
\end{equation}
where $a$ is some constant that depends on the spin moment. The ratio of the volume centered on the ground state, $\psi$ of the source Hamiltonian in Eq.~\ref{eq:SourceHamiltonian} to the volume flat space is then given by,
\begin{equation}
    \frac{V_{\psi}}{V_{\text{flat}}} \approx 1 - b R h^2 
\end{equation}
with $b$ a constant and $R$ the scalar curvature. 

In the XY chain, the Euler characteristic was examined for a manifold parameterized by the spin anisotropy and by a global rotation. The topology of the resulting manifold was shown to change depending on whether or not the model was in a paramagnetic or ordered phase, with the former phase resulting in an irrational Euler characteristic due to the emergence of a cusp like singular~\cite{kolodrubetz2017geometry}.

Another application of quantum curvature is in the context of variational quantum Monte Carlo, where the QFIM is used to find the energetic minimum using stochastic reconfiguration~\cite{sorella2001generalized}. In this case, the local curvature of the state space can be used to infer the robustness of the variational calculation, as discussed in Ref.~\cite{park2020geometry}.

\section{Summary and Outlook}\label{sec:Summary}

In this paper, we have presented a pedagogical introduction to information geometry. Though the formal foundations of information geometry were laid over a century ago, it is in the last two decades that we have seen a rapid growth in research interest in the many fields to which this geometry is applicable. In particular, there are now an abundance of studies examining Fidelity susceptibility and Fisher information in various models from condensed matter systems to cold atom setups. We expect that this interest will only grow.  Beyond equilibrium physics, concepts from quantum information geometry are finding application in the study of dynamical phases~\cite{mera2018dynamical} and quantum chaos~\cite{wang2011chaos}. 

There is still much to be learned about the geometrical structure of quantum state space and in particular about interpretations of geometric invariants such as the curvature as discussed in ~\cite{erdmenger2020information}. If one recalls Eq.~\ref{eq:PotentialFormOfMetric}, we see that in classical information geometry, the metric is sourced by the Shannon information. In quantum information geometry, the analogous quantity is the K\"{a}hler potential (see, for example, Ref.~\cite{bengtsson2017geometry}). Investigation of this quantity in many-body systems is only just beginning, and there are no doubt many exciting results to be found in this direction.

\appendix
\section{Properties of Shannon Information}
\label{app:PropertiesOfInformation}

In this appendix we discuss some properties of information measures. We use $P=\{p_1,...,p_N\}$ to denote a discrete distribution. The Shannon information defined in Eq.~\ref{eq:ShannonInformation} has the following properties~\cite{bengtsson2017geometry},
\begin{itemize}
    \item \emph{Positivity}, $S(P)\geq 0$ 
    \item \emph{Continuity}, $S(P)$ will vary continuous with the distribution $P$.
    \item \emph{Expansibility}, $S(p_1,...,p_N) = S(p_1,...,p_N,0)$.
    \item \emph{Concavity}, $S(\alpha P_1 + (1-\alpha)P_2)\geq \alpha S(P_1) + (1-\alpha) S(P_2)$. The mixing of distributions always increases the information.
    \item \emph{Subadditivity} For two random variables, not necessarily independent, where $p_{12}$ is the joint probability distribution,
    \begin{equation}
        S(P_{12}) \leq S(P_1) + S(P_2)
    \end{equation}
    with equality only if the two distributions are independent.
    \item \emph{Recursion}. We can coarse grain a distribution $P = {p_1,\ldots,p_N}$, via,
    \begin{equation}
        q_j = \sum_{i=k_{j-1}+1}^{k_j} p_i
    \end{equation}
    where $j=1,\ldots,r$ and $k_0=0$.
    The discrete distribution has been partition according to the sum $N=\sum_{i=1}^r k_i$. Then the Shannon information is given by, 
    \begin{equation}
        S(P) = S(Q) + q_1S\left(\frac{p_1}{q_1},\ldots,\frac{p_{k_1}}{q_1}\right)
               + \ldots + q_r S\left(\frac{p_{k_{N-k_r+1}}}{q_r},\ldots,\frac{p_N}{q_r}\right)
    \end{equation}
\end{itemize}
It is the property of recursion that makes the Shannon information unique amongst all possible choices of information measure. It appears in many other interesting contexts. In particular in the field of signal transmission where it was originally developed.

\section{Example Calculations}

\subsection{Single Qubit}\label{App:SampleCalculationSingleQubit}
We may compute the trace explicitly by diagonalizing the Hamiltonian with source
term. We denote the coefficients of the magnetization operator as $m_\alpha$ (no
hat) and temporarily suppress the explicit dependence on $\theta$ and $\phi$
\begin{equation}
  \hat{H} \doteq 
  \begin{pmatrix}
    \Delta-hm_z  &  h(m_x - im_y)   \\
    h(m_x + i m_y)  & -(\Delta-hm_z)
  \end{pmatrix} \nonumber
\end{equation}
We can immediately read off the eigenvalues for this matrix,
\begin{align}
  \lambda_{\pm} &= \pm\sqrt{(\Delta-hm_z)^2 + h^2(m_x-im_y)(m_x+im_y)} \nonumber \\
                &= \pm\sqrt{\Delta^2 - 2\Delta m_zh + h^2} \nonumber
\end{align}
and thus find,
\begin{equation}
  \mathcal{Z}(h) =  2\cosh(\beta\sqrt{\Delta^2 - 2\Delta m_z h + h^2})
\end{equation}
The variance is given by,
\begin{align}
  \text{Var}(\hat{M}) &= \tr(\rho\hat{M}^2) - \tr(\rho\hat{M})^2 \nonumber \\ 
  &= 1 - \tanh^2(\beta\qty|\Delta|)\cos^2(\phi) 
  \label{eq:totVar}
\end{align}
The thermal variance can be evaluated by first computing the first and second
derivatives of the partition function with respect to the driving field,
\begin{subequations}
  \begin{align}
    \partial_{h=0}\mathcal{Z} &=
    \partial_{h=0}2\cosh(\beta\sqrt{\Delta^2-2\Delta m_z h + h^2}) \nonumber \\
    &=2\beta\left(\frac{\sinh(\beta\sqrt{\Delta^2-2\Delta m_z h + h^2})}
                {\sqrt{\Delta^2 -2\Delta m_z h + h^2}}  \left(h - \Delta
                m_z\right)\right)_{h=0} \nonumber \\
                &=-2\beta\text{sgn}(\Delta) m_z\sinh(\beta\qty|\Delta|) \\
    \partial_{h=0}^2 \mathcal{Z} 
    &= 
    \left(
      2\beta^2 \frac{\cosh(\beta\sqrt{\Delta^2-2\Delta m_z h + h^2})}
                    {\Delta^2 - 2\Delta m_z h + h^2} 
                    (h-\Delta m_z)^2 \right. \nonumber \\
                   & \left.
      -2\beta\frac{\sinh(\beta\sqrt{\Delta^2 - 2\Delta m_z h + h^2})}
                  {\left(\Delta^2 - 2\Delta m_z h + h^2\right)^\frac{3}{2}}
                  (h-\Delta m_z)^2
    \right. \nonumber \\
    &\quad+ \left. 2\beta\frac{\sinh(\beta\sqrt{\Delta^2-2\Delta m_z h + h^2})}
    {\sqrt{\Delta^2 -2\Delta m_z h + h^2}}\right)_{h=0} \nonumber \\
    &= 2\beta^2 \cosh(\beta\qty|\Delta|)m_z^2
      -2\beta   \frac{\sinh(\beta\qty|\Delta|)}{|\Delta|}m_z^2
      +2\beta   \frac{\sinh(\beta\qty|\Delta|)}{\qty|\Delta|} \nonumber \\
    &= 2\beta^2\cosh(\beta\qty|\Delta|)m_z^2 +
    2\beta(1-m_z^2)\frac{\sinh(\beta|\Delta|)}{\qty|\Delta|}
  \end{align}
\end{subequations}
No we substitute these relations into the definition of the thermal variance
which we restate below,
\begin{align}
  \text{Var}_\mathcal{T}(\hat{M};\hat{\rho}) &= \frac{1}{\beta^2}
  \left(\frac{1}{\mathcal{Z}(h=0)}\left(\pdv[2]{\mathcal{Z}}{h}\right)_{h=0} 
  -\frac{1}{\mathcal{Z}^2(h=0)}\left(\pdv{\mathcal{Z}}{h}\right)_{h=0}^2
  \right) \nonumber \\
  &= \frac{1}{\beta^2}\left(\beta^2m_z^2 +
  \beta(1-m_z^2)\frac{\tanh(\beta\qty|\Delta|)}{\qty|\Delta|}
  -\beta^2m_z^2\tanh^2(\beta\qty|\Delta|)\right) \nonumber \\
  &= m_z^2 +(1-m_z^2)\frac{\tanh(\beta\qty|\Delta|)}{\beta\qty|\Delta|} -
  m_z^2\tanh^2(\beta\qty|\Delta|) \nonumber \\
  &= m_z^2(1-\tanh^2(\beta\qty|\Delta|)) +
  (1-m_z^2)\frac{\tanh(\beta\qty|\Delta|)}{\beta\qty|\Delta|}
  \label{eq:thermVar}
\end{align}
The QV can be computed by substituting Eq.~\ref{eq:totVar} and
Eq.~\ref{eq:thermVar} into Eq.~\ref{eq:QuantumCovariance},
\begin{align}
  \text{Var}_\mathcal{Q}(\hat{M};\hat{\rho}) &=
  1-\tanh^2(\beta\qty|\Delta|)m_z^2 - m_z^2(1-\tanh^2(\beta\qty|\Delta|)) 
  - (1-m_z^2)\frac{\tanh(\beta\qty|\Delta|)}{\beta\qty|\Delta|} \nonumber \\
  &= (1-m_z^2)\left(1 - \frac{\tanh(\beta\qty|\Delta|)}{\beta\qty|\Delta|}\right)
\end{align}
Using Eq.~\ref{eq:DefnQFIMUnitaryPart}, we can calculate an explicit form of
the QFI for the single qubit as well. Since diagonal entries don't contribute
to the QFI, we consider only off diagonal terms, and denote by
$\lambda_{\pm}=\pm\Delta$
the two eigenvalues and eigenstates of the Hamiltonian. We denote the
corresponding eigenstates of $\sigma^z$ by $\ket{\pm}$. The QFI given by,
\begin{align}
  \mathcal{F}(\hat{M};\hat{\rho}) &=
  2\sum_{\lambda,\lambda'}\frac{(p_\lambda-p_{\lambda'})^2}{p_\lambda+p_{\lambda'}}
  \qty|\bra{\lambda}\hat{M}\ket{\lambda'}|^2 \nonumber \\
  &= 
  4 \frac{(p_+-p_{-})^2}{p_+ +p_{-}}
  \qty|\bra{+}\hat{M}\ket{-}|^2  \nonumber \\
  &= \frac{4}{\mathcal{Z}}\frac{(e^{-\beta\qty|\Delta|} - e^{\beta\qty|\Delta|})^2}
           {e^{-\beta\qty|\Delta|} + e^{\beta\qty|\Delta|}} (m_x^2 + m_y^2) \nonumber \\
           &= \frac{4}{\mathcal{Z}}
     \frac{4\sinh^2(\beta\qty|\Delta|)}{2\cosh(\beta\qty|\Delta|)} (1-m_z^2)\nonumber \\
  &= 4\tanh^2(\beta\qty|\Delta|)(1-m_z^2) \nonumber \\
  &= 4\tanh^2(\beta\qty|\Delta|)\sin^2(\phi)
  \label{eq:qfiQubit}
\end{align}

\bibliographystyle{unsrt}
\bibliography{refs}

\end{document}